\documentclass{article}
\usepackage{amssymb}
\usepackage{multicol,multienum}
\usepackage{latexsym,amssymb,amsmath}
\usepackage{indentfirst}
\usepackage{cases}
\usepackage{color}
\usepackage{graphicx}
\usepackage{mathrsfs}
\usepackage{amsthm}
\usepackage{bm}
\usepackage{CJK}
\usepackage{url}
\usepackage[font={sf,bf},textfont=md]{caption}
\usepackage{animate}
\usepackage{hyperref}
\usepackage{multirow}
\usepackage{interval}
\usepackage[table,xcdraw]{xcolor}
\usepackage{subcaption}
\usepackage{caption}
\usepackage{bm}
\usepackage{thmtools, thm-restate}
\usepackage{bbm}
\usepackage{verbatim}
\usepackage{wrapfig}
\usepackage{lscape}
\usepackage{rotating}
\usepackage{algorithm}
\usepackage{enumitem}
\usepackage{natbib}
\usepackage[noend]{algpseudocode}
\usepackage{bbding}
\usepackage{booktabs}
\usepackage{titlesec}
\usepackage{bigstrut}
\usepackage[stable]{footmisc}
\usepackage{multibib}
\usepackage{pgffor}
\newcites{online}{References}

\newcommand{\irow}[1]{
  \big(\begin{smallmatrix}#1\end{smallmatrix}\big)%
}

\newcommand{\beginsupplement}{%
        \setcounter{table}{0}
        \renewcommand{\thetable}{S\arabic{table}}%
        \setcounter{figure}{0}
        \renewcommand{\thefigure}{S\arabic{figure}}%
        \setcounter{page}{0}
        \renewcommand{\theequation}{S.\arabic{equation}}%
        \setcounter{equation}{0}
     }

\newtheorem{theorem}{Theorem}
\newtheorem{lemma}{Lemma}

\newtheorem{assumption}{Assumption}

\begin{document}

\def\spacingset#1{\renewcommand{\baselinestretch}%
{#1}\small\normalsize} \spacingset{1}

\title{\bf Selection of Regression Models under Linear Restrictions for Fixed and Random Designs}
\author{Sen Tian\footnote{E-mail: st1864@stern.nyu.edu} \quad Clifford M. Hurvich  \quad Jeffrey S. Simonoff \\\\
  Department of Technology, Operations, and Statistics, \\Stern School of Business, New York University.}
\date{}
\maketitle

\begin{abstract}

Many important modeling tasks in linear regression, including variable selection (in which slopes of some predictors are set equal to zero) and simplified models based on sums or differences of predictors (in which slopes of those predictors are set equal to each other, or the negative of each other, respectively), can be viewed as being based on imposing linear restrictions on regression parameters. In this paper, we discuss how such models can be compared using information criteria designed to estimate predictive measures like squared error and Kullback-Leibler (KL) discrepancy, in the presence of either deterministic predictors (fixed-X) or random predictors (random-X). We extend the justifications for existing fixed-X criteria C$_p$, FPE and AICc, and random-X criteria S$_p$ and RC$_p$, to general linear restrictions. We further propose and justify a KL-based criterion, RAICc, under random-X for variable selection and general linear restrictions. We show in simulations that the use of the KL-based criteria AICc and RAICc results in better predictive performance and sparser solutions than the use of squared error-based criteria, including cross-validation. Supplemental material containing the technical details of the theorems is attached at the end of the main document. The computer code to reproduce the results for this article, and the complete set of simulation results, are available online\footnote{https://github.com/sentian/RAICc.}.
\end{abstract}

\noindent%
{\it Keywords:} AICc; C$_p$; Information criteria; Optimism; Random-X

\section{Introduction}
\subsection{Model selection under linear restrictions}
Consider a linear regression problem with an $n\times 1$ response vector $y$ and an $n\times p$ design matrix $X$. The true model is generated from 
\begin{equation}
y=X \beta_0 + \epsilon,
\label{eq:truemodel}
\end{equation}
where $\beta_0$ is a $p \times 1$ true coefficient vector, and the $n \times 1$ vector $\epsilon$ is independent of $X$, with $\{\epsilon_i\}_{i=1}^n \stackrel {iid} {\sim} N(0,\sigma_0^2)$. Note that $\beta_0$ represents the true parameters, not an intercept term. 
We consider an approximating model 
\begin{equation*}
y=X \beta + u,
\end{equation*}
where $\beta$ is $p \times 1$ and the $n \times 1$ vector $u$ is independent of $X$, with $\{u_i\}_{i=1}^n \stackrel {iid} {\sim} N(0,\sigma^2)$. For this approximating model, we further impose $m$ linear restrictions on the coefficient vectors $\beta$ that are given by
\begin{equation}
  R \beta = r,
  \label{eq:restriction}
\end{equation}
where $R$ is an $m \times p$ matrix with linearly independent rows ($\text{rank}(R)=m$) and $r$ is an $m \times 1$ vector. Both $R$ and $r$ are nonrandom. Examples of such restrictions include setting some slopes equal to 0 (which corresponds to variable selection), setting slopes equal to each other (which corresponds to using the sum of predictors in a model), and setting sums of slopes to 0 (which for pairs of predictors corresponds to using the difference of the predictors in a model).

Suppose first that $X$ is deterministic; we refer to this as the fixed-X design. Denote $f(y_i|x_i,\beta,\sigma^2)$ as the density for $y$ conditional on the $i$-th row of $x_i$. We have the log-likelihood function (multiplied by $-2$)
\begin{equation}
-2 \log f(y|X,\beta,\sigma^2) = -2 \sum_{i=1}^n \log f(y_i|x_i,\beta,\sigma^2) = n \log (2\pi \sigma^2) + \frac{1}{\sigma^2} || y-X\beta||_2^2.
\label{eq:loglike_fixedx}
\end{equation}
By minimizing \eqref{eq:loglike_fixedx} subject to \eqref{eq:restriction}, we obtain the restricted maximum likelihood estimator (MLE)
\begin{equation}
\begin{aligned}
\hat{\beta} &= \hat{\beta}^f + (X^T X)^{-1} R^T ( R(X^T X)^{-1} R^T)^{-1} (r-R \hat{\beta}^f),\\ 
\hat \sigma^2 &= \frac{1}{n} ||y-X \hat{\beta}||^2, 
\end{aligned}
\label{eq:betahat_sigmahatsq}
\end{equation}
where $\hat{\beta}^f = (X^T X)^{-1} X^T y$ is the unrestricted least squares estimator. Since the errors are assumed to be Gaussian, $\hat\beta$ is also the restricted least squares estimator. 

In practice, a sequence of estimators $\hat\beta(R_i,r_i|X,y)$, each based on a different set of restrictions, is often generated, and the goal is to choose the one with the best predictive performance. This can be done on the basis of information criteria, which are designed to estimate the predictive accuracy for each considered model. Note that the notion of predictive accuracy can be as simple as distance of a predicted value from a future value, as is the case in squared-error prediction measures, but also can encompass the more general idea that the log-likelihood is a measure of the accuracy of a fitted distribution as a prediction for the distribution of a future observation. This idea can be traced back to \citet{Akaike1973}, as noted in an interview with Akaike \citep{findley1995conversation}; see also \citet{efron1986biased}. 

\subsection{Variable selection under fixed-X}
\label{sec:intro_subsetselection}
An important example of comparing models with different linear restrictions on $\beta$ is variable selection. We consider fitting the ordinary least squares (OLS) estimator on a predetermined subset of predictors with size $k$, and without loss of generality, the subset includes the first $k$ predictors of $X$, i.e. $\hat\beta^f(X_1,\cdots,X_k,y)$. By letting $R_k= \irow{0 & I_{p-k}}_{(p-k) \times p} $ and $r_k = \irow{0}_{(p-k) \times 1}$, it is easy to verify that $\hat{\beta}(R_k,r_k|X,y)=\hat\beta^f(X_1,\cdots,X_k,y)$. Therefore, comparing OLS fits on different subsets of predictors falls into the framework of comparing estimators with different linear restrictions on $\beta$.

Information criteria are designed to provide an unbiased estimate of the test error. We simplify the notation by denoting $\hat\beta(k) = \hat{\beta}(R_k,r_k|X,y)$. We also denote errF as the in-sample training error and ErrF as the out-of-sample test error. errF measures how well the estimated model fits on the training data $(X,y)$, while ErrF measures how well the estimated model predicts the new test data $(X,\tilde{y})$, where $\tilde{y}$ is an independent copy of the original response $y$, i.e. $\tilde{y}$ is drawn from the conditional distribution of $y|X$. The notations of errF and ErrF are based on those in \citet{efron2004estimation}, and the notation F here indicates that we have a fixed-$X$ design. \citet{efron1986biased} defined the optimism of a fitting procedure as the difference between the test error and the training error, i.e.
\begin{equation*}
\text{optF} = \text{ErrF} - \text{errF},
\end{equation*}
and introduced the optimism theorem,
\begin{equation*}
E_y(\text{optF}) = E_y(\text{ErrF}) - E_y(\text{errF}),
\end{equation*}
where $E_y$ represents the expectation taken under the true model with respect to the random variable $y$. The optimism theorem provides an elegant framework to obtain an unbiased estimator of $E(\text{ErrF})$, 
that is
\begin{equation*}
\widehat{\text{ErrF}} = \text{errF} + E_y(\text{optF}),
\end{equation*}
where the notation $\widehat{\text{ErrF}}$ follows from \citet{efron2004estimation}. It turns out that many existing information criteria can be derived using the concept of optimism. 

A typical measure of the discrepancy between the true model and an approximating model is the squared error (SE), i.e. 
\begin{equation*}
\text{ErrF}_\text{SE} = E_{\tilde{y}}\left( \lVert \tilde{y}-X\hat{\beta} \rVert_2^2 \right).
\end{equation*} 
The training error is $\text{errF}_\text{SE} = \displaystyle \lVert y-X\hat{\beta} \rVert_2^2$. \citet{ye1998measuring} and \citet{efron2004estimation} showed that for any general fitting procedure $\hat\mu$ and any model distribution (not necessarily Gaussian)
\begin{equation}
E_y(\text{optF}_\text{SE}) = 2\sum_{i=1}^n \text{Cov}_y(\hat\mu_i,y_i),
\label{eq:EoptF_SE}
\end{equation}
which is often referred to as the covariance penalty. For the OLS estimator $\hat\mu(k) = X\hat\beta(k)$ it is easy to verify that $E_y(\text{optF}_\text{SE}(k)) = 2 \sigma_0^2 k$. We denote RSS$(k)$ as the residual sum of squares for the OLS estimator, i.e. $\text{RSS}(k)=\lVert y- X\hat\beta(k) \rVert_2^2$. Hence,
\begin{equation*}
\widehat{\text{ErrF}}_\text{SE}(k) = \text{RSS}(k) + 2 \sigma_0^2 k
\end{equation*}
is an unbiased estimator of $E_y(\text{ErrF}_\text{SE})$. As suggested by \citet{mallows1973some}, typically $\sigma_0^2$ is estimated using the OLS fit on all the predictors, i.e. $\hat\sigma_0^2=\text{RSS}(p)/(n-p)$. We then obtain the Mallows' C$_p$ criterion \citep{mallows1973some} 
\begin{equation}
\text{C}_p(k) = \text{RSS}(k) + \frac{\text{RSS}(p)}{n-p} 2k.
\label{eq:cp_subsetselection}
\end{equation}
An alternative is to use the OLS fit based on the $k$ predictors in the subset to estimate $\sigma_0^2$. i.e. $\hat\sigma_0^2 = \text{RSS}(k)/(n-k)$, which yields the final prediction error \citep{akaike1969fitting,akaike1970statistical}
\begin{equation}
\text{FPE}(k) = \text{RSS}(k)\frac{n+k}{n-k}.
\label{eq:cptilde_subsetselection}
\end{equation}

Another commonly-used error measure is (twice) the Kullback-Leibler (KL) divergence \citep[see, e.g.,][Section 3]{konishi2008information}
\begin{equation}
\text{KLF} = E_{\tilde{y}}\left[ 2\log f(\tilde{y} | X,\beta_0,\sigma_0^2 ) - 2\log f(\tilde{y} | X,\hat\beta,\hat\sigma^2 ) \right].
\label{eq:KLF}
\end{equation}
The right-hand side of \eqref{eq:KLF} evaluates the predictive accuracy of the fitted model, by measuring the closeness of the distribution of $\tilde{y}$ based on the fitted model and the distribution of $\tilde{y}$ based on the true model. The term $E_{\tilde{y}}\left[ 2\log f(\tilde{y} | X,\beta_0,\sigma_0^2 ) \right]$ is the same for every fitted model. Therefore, an equivalent error measure is the expected likelihood
\begin{equation*}
\text{ErrF}_\text{KL} = E_{\tilde{y}}\left[ -2\log f(\tilde{y} | X,\hat\beta,\hat\sigma^2 ) \right].
\end{equation*}
The training error is 
\begin{equation*}
\text{errF}_\text{KL} = -2\log f(y|X,\hat\beta,\hat\sigma^2).
\end{equation*}
For the OLS estimator $\hat\beta(k)$, \citet{sugiura1978further} and \citet{hurvich1989regression} showed that under the Gaussian error \eqref{eq:truemodel}
\begin{equation*}
E_y(\text{optF}_\text{KL}(k)) = n\frac{n+k}{n-k-2}-n,
\end{equation*}
and hence
\begin{equation*}
\widehat{\text{ErrF}}_\text{KL}(k) = n\log\left(\frac{\text{RSS}(k)}{n}\right) + n\frac{n+k}{n-k-2} + n\log(2\pi)
\end{equation*}
is an unbiased estimator of $E_y(\text{ErrF}_\text{KL})$. Since the term $n\log(2\pi)$ appears in all of the models being compared, and thus is irrelevant when comparing criteria for the models, the authors dropped it and introduced the corrected AIC
\begin{equation}
\text{AICc}(k) = n \log\left( \frac{\text{RSS}(k)}{n}\right) + n\frac{n+k}{n-k-2}.
\label{eq:aicc_subsetselection}
\end{equation}
\citet{Hurvich1991} showed that AICc has superior finite-sample predictive performance compared to AIC \citep{Akaike1973}
\begin{equation*}
\text{AIC}(k) = n \log\left( \frac{\text{RSS}(k)}{n}\right) + n + 2(k+1),
\end{equation*}
which does not require a Gaussian error assumption but relies on asymptotic results. The derivations of AICc and AIC require the assumption that the true model is included in the approximating models. Neither AICc nor AIC involve $\sigma_0^2$, a clear advantage over C$_p$. Note that the second term of AICc can be rewritten as $n[1 + (2k+2)/(n-k-2)]$, which approximately equals the sum of the second and third terms of AIC when $n$ is large relative to $k$, demonstrating their asymptotic equivalence when $n\rightarrow\infty$ and $p$ is fixed.

\subsection{From fixed-X to random-X}
The assumption that $X$ is fixed holds in many applications, for example in a designed experiment where categorical predictors are represented using indicator variables or effect codings. However, in many other cases where the data are observational and the experiment is conducted in an uncontrolled manner, fixed-X is not valid and it is more appropriate to treat $(x_i,y_i)_{i=1}^n$ as $iid$ random draws from the joint distribution of $X$ and $y$. We refer to this as the random-X design. 

As noted by \citet{breiman1992submodel}, the choice between fixed-X and random-X is conceptual, and is normally determined based on the nature of the study. The extra source of randomness from $X$ results in larger test error compared to the fixed-X situation, and therefore the information criteria designed under fixed-X can be biased estimates of the random-X test error. Furthermore, when applied as selection rules, the authors showed in simulations that C$_p$ leads to significant overfitting under the random-X design. This motivates the derivation of information criteria for the random-X situation. 

For the random-X design, we assume that the row vectors of $X$, $\{x_i\}_{i=1}^n$, are $iid$ multivariate normal with mean $E(x_i)=0$ and covariance matrix $E(x_i x_i^T)=\Sigma_0$. Let $f(y_i,x_i|\beta,\sigma^2,\Sigma)$ denote the joint multivariate normal density for $y_i$ and $x_i$. Let $g(x_i|\Sigma)$ denote the multivariate normal density for $x_i$. By partitioning the joint density of $(y,X)$ into the product of the conditional and marginal densities, and by separating the parameters of interest, we have the log-likelihood function (multiplied by $-2$)
\begin{equation}
\begin{aligned}
-2 \log f(y, X|\beta,\sigma^2,\Sigma) &= \sum_{i=1}^n -2 \log f(y_i, x_i|\beta,\sigma^2,\Sigma) = -2 \sum_{i=1}^n [\log f(y_i|x_i,\beta,\sigma^2) + \log g(x_i|\Sigma)] \\
&= \left [ n \log (2\pi \sigma^2) + \frac{1}{\sigma^2} || y-X\beta||_2^2 \right ] + \left [np \log(2\pi) + n \log |\Sigma| + \sum_{i=1}^n x_i^T \Sigma^{-1} x_i \right ].
\end{aligned}
\label{eq:loglike_randomx}
\end{equation}
Minimizing \eqref{eq:loglike_randomx} subject to \eqref{eq:restriction}, we find that the MLE $(\hat{\beta}, \hat{\sigma}^2)$ of $(\beta, \sigma^2)$ remains the same as in the fixed-X design, i.e. \eqref{eq:betahat_sigmahatsq}. The MLE of $\Sigma$ is given by
\begin{equation*}
\hat \Sigma = \frac{1}{n} \sum_{i=1}^n x_i \, x_i^T = \frac{1}{n} X^T X.
\end{equation*}

Since $\hat\beta$ is unchanged when we move from fixed-X to random-X, variable selection as an example of linear restrictions on $\beta$ is based on the same parameter estimates as in Section \ref{sec:intro_subsetselection}. Denote errR, ErrR and optR as the training error, test error and the optimism under random-X, respectively. We generate $X^{(n)}$ as an independent copy of $X$, where the rows $\{x_i^{(n)}\}_{i=1}^n$ are $iid$ multivariate normal $\mathcal{N}(0,\Sigma_0)$. The new copy of the response $y^{(n)}$ is generated from the conditional distribution $y|X^{(n)}$. The optimism for random-X can be defined in the same way as for fixed-X, i.e. $\text{optR}=\text{ErrR}-\text{errR}$. \citet{rosset2020fixed} discussed the optimism for general fitting procedures, when the discrepancy between the true and approximating models is measured by the squared error (SE), i.e.
\begin{equation*}
\text{ErrR}_\text{SE} = E_{X^{(n)},y^{(n)}} \left( \lVert y^{(n)} - X^{(n)} \hat\beta \rVert_2^2 \right).
\end{equation*}
The training error is $\text{errR}_\text{SE} = \lVert y - X \hat\beta \rVert_2^2$. For the OLS estimator, the authors showed that 
\begin{equation*}
E_{X,y}(\text{optR}_\text{SE}(k)) = \sigma_0^2 k \left(2 + \frac{k+1}{n-k-1} \right),
\end{equation*}
and hence
\begin{equation*}
\widehat{\text{ErrR}}_\text{SE}(k) = \text{RSS}(k) + \sigma_0^2 k \left(2 + \frac{k+1}{n-k-1} \right)
\end{equation*}
is an unbiased estimator of $E_{X,y}(\text{ErrR}_\text{SE})$. The result holds for arbitrary joint distributions of $(x_y,y_i)$ , and it only requires $x_i$ being marginally normal. As in the fixed-X case, if we use the unbiased estimate of $\sigma_0^2$ based on the full OLS fit, we have the analog of the C$_p$ rule for random-X, 
\begin{equation}
\text{RC}_p(k) = \text{RSS}(k) + \frac{\text{RSS}(p)}{n-p} k\left(2 + \frac{k+1}{n-k-1}\right).
\label{eq:rcp_subsetselection}
\end{equation}
If we use the alternative estimate of $\sigma_0^2$ based on the OLS fit on the $k$ predictors in the subset, i.e. $\hat\sigma_0^2=\text{RSS}(k)/(n-k)$, we have the analog of the FPE rule for random-X,
\begin{equation}
\text{S}_p(k) = \text{RSS}(k)\frac{n(n-1)}{(n-k)(n-k-1)}.
\label{eq:sp_subsetselection}
\end{equation}
\citet{hocking1976biometrics} refers to \eqref{eq:sp_subsetselection} as the S$_p$ criterion of \citet{sclove1969criteria}; see also \citet{thompson1978a,thompson1978b}. Note that the notation used here is slightly different from that in \citet{rosset2020fixed}, where the authors used RC$_p$ to denote the infeasible criterion involving $\sigma_0^2$ and used $\widehat{\text{RC}}_p$ to denote the feasible criterion S$_p$. The RC$_p$ criterion in our notation was not studied in their paper. 

Another class of selection rules is cross-validation (CV), which does not impose parametric assumptions on the model. A commonly used type of CV is the so-called K-fold CV. The data are randomly split into K equal folds. For each fold, the model is fitted using data in the remaining folds and is evaluated on the current fold. The process is repeated for all K folds, and an average squared error is obtained. In particular, the n-fold CV or leave-one-out (LOO) CV provides an approximately unbiased estimator of the test error under the random-X design, i.e. $E_{X,y}(\text{ErrR}_\text{SE})$. \citet{burman1989comparative} showed that for OLS, LOOCV has the smallest bias and variance in estimating the squared error-based test error, among all K-fold CV estimators. LOOCV is generally not preferred due to its large computational cost, but for OLS, the LOOCV error estimate has an analytical expression: the predicted residual sum of squares (PRESS) statistic \citep{allen1974relationship}
\begin{equation*}
\text{PRESS}(k) = \sum_{i=1}^n \left( \frac{y_i-x_i^T\hat\beta(k)}{1-H_{ii}(k)} \right)^2,
\end{equation*}
where $H(k) = X(k)(X(k)^T X(k))^{-1}X(k)^T$ and $X(k)$ contains the first $k$ columns of $X$. 

\subsection{General linear restrictions}
Variable selection is a special case of linear restrictions on $\beta$, where certain entries of $\beta$ are restricted to be zero. In practice, we may restrict predictors to have the same coefficient (e.g. $\beta_1=\beta_2=\beta_3$), or we may restrict the sum of their effects (e.g. $\beta_1+\beta_2+\beta_3=1$). Using the structure in \eqref{eq:restriction}, we formulate a sequence of models, each of which imposes a set of general restrictions on $\beta$, where the goal is to select the model with best predictive performance. The previously defined information criteria and PRESS cannot be applied to this problem, although \citet{tarpey2000note} derived the PRESS statistic for the estimator under general restrictions as
\begin{equation*}
\text{PRESS}(R,r) = \sum_{i=1}^n \left( \frac{y_i-x_i^T\hat\beta}{1-H_{ii}+{H_Q}_{ii}} \right)^2,
\end{equation*}
where $H=X(X^T X)^{-1} X^T$ and $H_Q = X (X^T X)^{-1} R^T \left[ R (X^T X)^{-1} R^T \right]^{-1} R (X^T X)^{-1} X^T$.

\subsection{The contribution of this paper}
The information criteria introduced in Section \ref{sec:intro_subsetselection} have been studied primarily in the context of variable selection problems under fixed-X. In this paper we discuss how such criteria can be generalized to model comparison under general linear restrictions with either a fixed-X or a random-X (in both cases including the special case of variable selection). Note that a selection rule is preferred if it chooses the models that lead to the best predictive performance. This is related to, but not the same as, providing the best estimate of the test error. These two goals are fundamentally different \citep[see, e.g.,][Section 7]{hastie2009elements}, and we focus on the predictive performance of the selected model.

In Section \ref{sec:ic_fixedx}, we consider the fixed-X situation and derive general versions of AICc, C$_p$ and FPE for arbitrary linear restrictions on $\beta$. Random-X is assumed in Section \ref{sec:ic_randomx} and a version of RC$_p$ and S$_p$ for general linear restrictions is obtained. Furthermore, we propose and justify a novel criterion, RAICc, for general linear restrictions and discuss its connections with AICc. We further show that expressions of the information criteria for variable selection problems can be recovered as special cases of their expressions derived under general restrictions. In Section \ref{sec:simulation}, we show via simulations that AICc and RAICc provide consistently strong predictive performance for both variable selection and general restriction problems. Lastly, in Section \ref{sec:conclusion}, we provide conclusions and discussions of potential future work.

\section{ Information criteria for fixed-X }
\label{sec:ic_fixedx}
\subsection{KL-based information criterion}

Using the likelihood function \eqref{eq:loglike_fixedx} and the MLE \eqref{eq:betahat_sigmahatsq}, the expected log-likelihood can be derived as 
\begin{equation*}
\begin{aligned}
\text{ErrF}_\text{KL} &=  E_{\tilde{y}} [-2 \log f( \tilde{y} | X,\hat\beta,\hat\sigma^2 )] =  n \log (2\pi \hat\sigma^2) + \frac{1}{\hat\sigma^2} E_{\tilde{y}} || \tilde{y}-X\hat\beta||_2^2 \\
&= n \log (2\pi \hat\sigma^2) + \frac{1}{\hat\sigma^2}  (\hat\beta-\beta_0)^T X^T X (\hat\beta-\beta_0) + \frac{n\sigma_0^2}{\hat\sigma^2},
\end{aligned}
\end{equation*}
and the training error is 
\begin{equation*}
\text{errF}_\text{KL} = -2\log f(y|X,\hat\beta,\hat\sigma^2) = n\log(2\pi\hat\sigma^2) + n.
\end{equation*}
In the context of variable selection, the assumption that the approximating model includes the true model is used in the derivations of AIC \citep{linhart1986model} and AICc \citep{hurvich1989regression}. This assumption can be generalized to the context of general restrictions. 
\begin{assumption}
If the approximating model satisfies the restrictions $R\beta = r$, then the true model satisfies the analogous restrictions $R\beta_0 = r$; that is, the true model is at least as restrictive as the approximating model. 
\label{assumption}
\end{assumption}
Under this assumption, we have the following lemma. The proofs for all of the lemmas and theorems in this paper are given in the Supplemental Material.   
\begin{lemma}
  Under Assumption \ref{assumption}, $\hat\sigma^2$ and the quadratic form $(\hat \beta-\beta_0)^T X^T X (\hat \beta-\beta_0)$ are independent, and 
  \begin{equation*}
  \begin{aligned}
    n \sigma_0^2 E_y\left[ \frac{1}{\hat{\sigma}^2} \right] &= n\frac{n}{n-p+m-2},\\
    E_y  \left [ (\hat \beta-\beta_0)^T X^T X (\hat \beta-\beta_0) \right ] &= \sigma_0^2 (p-m).
  \end{aligned}
  \end{equation*}
\label{thm:components_ekl_lr_fixedx}
\end{lemma}
Lemma \ref{thm:components_ekl_lr_fixedx} provides the fundamentals for calculating the expected optimism.
\begin{theorem}
Under Assumption \ref{assumption}, 
\begin{equation*}
E_y(\text{optF}_\text{KL}) = n \frac{n+p-m}{n-p+m-2} - n.
\end{equation*}
\label{thm:EoptF_KL}
\end{theorem}

Consequently, 
\begin{equation*}
\widehat{\text{ErrF}}_\text{KL} = \text{errF}_\text{KL} + E_y(\text{optF}_\text{KL}) = n\log(\hat\sigma^2) + n \frac{n+p-m}{n-p+m-2} + n\log(2\pi)
\end{equation*}
is an unbiased estimator of the test error $E_y \left[ \text{ErrF}_\text{KL} \right]$. We follow the same tradition as in the derivations of AIC and AICc that since the term $n\log(2\pi)$ appears in $\widehat{\text{ErrF}}_\text{KL}$ for every model being compared, it is irrelevant for purposes of model selection. We therefore ignore this term and define 
\begin{equation*}
 \text{AICc}(R,r) = n\log \left( \frac{\text{RSS}(R,r)}{n}  \right) + n \frac{n+p-m}{n-p+m-2},
\end{equation*}
where RSS$(R,r)= \lVert y -X\hat\beta \rVert_2^2$. For the variable selection problem, e.g. regressing on a subset of predictors with size $k$, we are restricting $p-k$ slope coefficients to be zero. By plugging $\hat\beta = \hat\beta(k)$ and $m=p-k$ into the expressions of AICc$(R,r)$, we obtain AICc$(k)$ given in \eqref{eq:aicc_subsetselection}.

\subsection{Squared error-based information criterion}
The covariance penalty \eqref{eq:EoptF_SE} is defined for any general fitting procedure. By explicitly calculating the covariance term for $\hat\mu=X\hat\beta$, we can obtain the expected optimism.
\begin{theorem}
\begin{equation*}
E_y (\text{optF}_\text{SE}) = 2 \sigma_0^2 (p-m).
\end{equation*}
\label{thm:EoptF_SE}
\end{theorem}

An immediate consequence of this is that
\begin{equation*}
\widehat{\text{ErrF}}_\text{SE} = \text{errF}_\text{SE} + E_y(\text{optF}_\text{SE}) = \text{RSS}(R,r) + 2 \sigma_0^2 (p-m)
\end{equation*}
is an unbiased estimator of $E_y(\text{ErrF}_\text{SE})$. Using the unbiased estimator of $\sigma_0^2$ given by the OLS fit based on all of the predictors, i.e. $\hat\sigma_0^2=\text{RSS}(p)/(n-p)$, we define
\begin{equation*}
\text{C}_p(R,r) = \text{RSS}(R,r) + \frac{\text{RSS}(p)}{n-p} 2(p-m).
\end{equation*}
An alternative estimate of $\sigma_0^2$ is $\text{RSS}(R,r)/(n-p+m)$, which yields 
\begin{equation*}
\text{FPE}(R,r) = \text{RSS}(R,r)\frac{n+p-m}{n-p+m}.
\end{equation*}
For the variable selection problem, by substituting $m=p-k$ into the expressions of C$_p$ and FPE, we obtain the previously-noted definitions of them, i.e. C$_p$(k) and FPE(k) given in \eqref{eq:cp_subsetselection} and \eqref{eq:cptilde_subsetselection}, respectively. 

\section{ Information criteria for random-X }
\label{sec:ic_randomx}

\subsection{KL-based information criterion, RAICc}
We replace the unknown parameters by their MLE, and have the fitted model $f(\cdot|\hat\beta,\hat\sigma^2,\hat\Sigma)$. The KL information measures how well the fitted model predicts the new set of data $(X^{(n)},y^{(n)})$, in terms of the closeness of the distributions of $(X^{(n)},y^{(n)})$ based on the fitted model and the true model, i.e. 
\begin{equation}
\text{KLR} = E_{X^{(n)},y^{(n)}} \left[ 2\log f(X^{(n)},y^{(n)}|\beta_0,\sigma_0^2,\Sigma_0) -2 \log f(X^{(n)},y^{(n)}|\hat\beta,\hat\sigma^2,\hat\Sigma) \right].
\label{eq:KLR}
\end{equation}
An equivalent form for model comparisons is the expected log-likelihood
\begin{equation*}
\begin{aligned}
&\text{ErrR}_\text{KL} =  E_{X^{(n)},y^{(n)}} \left[ -2 \log f(X^{(n)},y^{(n)}|\hat\beta,\hat\sigma^2,\hat\Sigma) \right] \\
&= \left[ n \log (2\pi \hat\sigma^2) + \frac{1}{\hat\sigma^2} E_{X^{(n)},y^{(n)}} || y^{(n)}-X^{(n)}\hat\beta||_2^2 \right ] + \left [np \log(2\pi) + n \log |\hat\Sigma| + E_{X^{(n)}} \left(\sum_{i=1}^n {x_{i}^{(n)}}^T \hat\Sigma^{-1} x_{i}^{(n)} \right) \right]\\
&= \left[ n \log (2\pi \hat\sigma^2) + \frac{n}{\hat\sigma^2}  (\hat\beta-\beta_0)^T \Sigma_0 (\hat\beta-\beta_0) + \frac{n\sigma_0^2}{\hat\sigma^2} \right ] + \left [np \log(2\pi) + n \log |\hat\Sigma| + n \text{Tr}(\hat\Sigma^{-1}\Sigma_{0})\right],
\end{aligned}
\end{equation*}
and the training error is
\begin{equation*}
\text{errR}_\text{KL} = -2\log f(X,y|\hat\beta,\hat\sigma^2,\hat\Sigma) = \left [ n \log (2\pi \hat\sigma^2) + n \right ] + \left [np \log(2\pi) + n \log |\hat\Sigma| + np \right ].
\end{equation*}

As in the fixed-X case, we assume that the true model satisfies the restrictions, i.e. $R\beta_0=r$, and we obtain the following lemma.
\begin{lemma}
Under Assumption \ref{assumption}, $\hat\sigma^2$ and $(\hat{\beta}-\beta_0)^T \Sigma_0 (\hat{\beta}-\beta_0)$ are independent conditionally on $X$, and 
\begin{equation*}
\begin{aligned}
E \left[ \text{Tr}(\hat \Sigma^{-1}\Sigma_0) \right] &= \frac{np}{n-p-1},\\
E_{X,y}  \left [ (\hat \beta-\beta_0)^T \Sigma_0 (\hat \beta-\beta_0) \right ] &= \sigma_0^2 \frac{p-m}{n-p+m-1}.
\end{aligned}
\end{equation*}
\label{thm:components_ekl_lr_randomx}
\end{lemma}
Lemma \ref{thm:components_ekl_lr_randomx} provides the components for calculating the expected optimism.
\begin{theorem}
Under Assumption \ref{assumption}, 
\begin{equation*}
E_{X,y}(\text{optR}_\text{KL}) = n \frac{n(n-1)}{(n-p+m-2)(n-p+m-1)} + n \frac{np}{n-p-1} - n(p+1).
\end{equation*}
\label{thm:EoptR_KL}
\end{theorem}

Consequently, 
\begin{equation*}
\begin{aligned}
\widehat{\text{ErrR}}_\text{KL} &= \text{errR}_\text{KL} + E_{X,y} (\text{optR}_\text{KL}) \\
&=n\log \left(\hat\sigma^2\right) + n \frac{n(n-1)}{(n-p+m-2)(n-p+m-1)} + n\log(2\pi)(p+1) + n\frac{np}{n-p-1} + n\log|\hat\Sigma|
\end{aligned}
\end{equation*}
is an unbiased estimator of the test error $E_{X,y}(\text{ErrR}_\text{KL})$. Note that the last three terms are free of the restrictions and only depend on $n$, $p$ and $X$. They are the same when we compare two models with different restrictions on $\beta$, and are thus irrelevant when comparing criteria for any two such models. Therefore, for the purpose of model selection, we define
\begin{equation*}
\text{RAICc}(R,r) = n\log \left(\frac{\text{RSS}(R,r)}{n}\right) + n \frac{n(n-1)}{(n-p+m-2)(n-p+m-1)}.
\end{equation*}
An equivalent form is
\begin{equation*}
\text{RAICc}(R,r) = \text{AICc}(R,r) + \frac{n(p-m)(p-m+1)}{(n-p+m-1)(n-p+m-2)}.
\end{equation*} 
For linear regression on a subset of predictors with size $k$, we are restricting $p-k$ coefficients to be zero. By substituting $m=p-k$ and $\hat\beta = \hat\beta(k)$ into the expression of $\text{RAICc}(R,r)$, we obtain the RAICc criterion for the variable selection problem, i.e.
\begin{equation*}
\text{RAICc}(k) = n\log \left(\frac{\text{RSS}(k)}{n}\right) + n \frac{n(n-1)}{(n-k-2)(n-k-1)}.
\end{equation*}

\subsection{Squared error-based information criteria}
According to \citet[formula 6 and proposition 1]{rosset2020fixed}, $E_{X,y}(\text{optR}_\text{SE})$ can be decomposed into $E_{X,y}(\text{optF}_\text{SE})$ plus an excess bias term and an excess variance term. We calculate both terms for our estimator $\hat\beta$ and obtain the following theorem.
\begin{theorem}
Under Assumption \ref{assumption},
\begin{equation*}
E_{X,y}(\text{optR}_\text{SE}) = \sigma_0^2(p-m) \left( 2+ \frac{p-m+1}{n-p+m-1} \right).
\end{equation*}
\label{thm:EoptR_SE}
\end{theorem}
An immediate consequence is that
\begin{equation*}
\widehat{\text{ErrR}}_\text{SE} = \text{errR}_\text{SE} + E_{X,y} (\text{optR}_\text{SE}) = \text{RSS}(R,r) + \sigma_0^2(p-m) \left( 2+ \frac{p-m+1}{n-p+m-1} \right)
\end{equation*}
is an unbiased estimator of $E_{X,y}(\text{ErrR}_\text{SE})$. Using the OLS fit on all of the predictors to estimate $\sigma_0^2$, we have 
\begin{equation*}
\text{RC}_p(R,r) = \text{RSS}(R,r) + \frac{\text{RSS(p)}}{n-p}(p-m) \left(2+\frac{p-m+1}{n-p+m-1}\right).
\end{equation*}
An alternative estimate of $\sigma_0^2$ is $\text{RSS}(R,r)/(n-p+m)$, which yields 
\begin{equation*}
\text{S}_p(R,r) = \text{RSS}(R,r)\frac{n(n-1)}{(n-p+m)(n-p+m-1)}.
\end{equation*}
For the variable selection problem, by substituting $m=p-k$ into the expressions of RC$_p$ and S$_p$, we obtain the previously-noted definitions of them, i.e. RC$_p$(k) and S$_p$(k) given in \eqref{eq:rcp_subsetselection} and \eqref{eq:sp_subsetselection}, respectively. 

\section{Performance of the selectors}
\label{sec:simulation}
\subsection{Some other selectors}
In this section we use computer simulations to explore the behavior of different criteria when used for model selection under linear restrictions (variable selection and general linear restrictions). In addition to the criteria already discussed, we also consider two other well-known criteria: 
\begin{equation*}
\text{BIC}(k) = n \log\left( \frac{\text{RSS}(k)}{n} \right) +\log(n) k
\end{equation*}
\citep{schwarz1978estimating}, and generalized cross-validation (GCV)
\begin{equation}
\text{GCV}(k) = \text{RSS}(k)\frac{n^2}{(n-k)^2}.
\label{eq:gcv_subsetselection}
\end{equation}
BIC is a consistent criterion, in the sense that under some conditions, if the true model is among the candidate models, the probability of selecting the true model approaches one, as the sample size becomes infinite. GCV, derived by \citet{craven1978smoothing} in the context of smoothing, is equivalent to the mean square over degrees of freedom criterion proposed by \citet{tukey1967discussion}. By comparing the expressions of \eqref{eq:gcv_subsetselection} and \eqref{eq:sp_subsetselection}, GCV and S$_p$ only differ by a multiplicative factor of $1 + k / [(n-k)(n-1)]$.

By analogy to the criteria discussed in Section \ref{sec:ic_fixedx} and \ref{sec:ic_randomx}, if we substitute $k=p-m$ into the expressions of BIC$(k)$ and GCV$(k)$, we obtain their corresponding expressions for general linear restrictions BIC$(R,r)$ and GCV$(R,r)$, respectively. We also consider two types of the cross-validation (CV): 10-fold CV (denoted as 10FCV) and leave-one-out CV (LOOCV). The LOOCV is based on the PRESS(k) statistic for the variable selection problem and PRESS(R,r) for the general restriction problem.

\subsection{Random-X}
We first consider the variable selection problem. The candidate models include the predictors of $X$ in a nested fashion, i.e. the candidate model of size $k$ includes the first $k$ columns of $X$ ($X_1,\cdots,X_k$). We describe the simulation settings reported here; description of and results from all other settings (243 configurations in total) can be found in the Online Supplemental Material\footnote{\url{https://github.com/sentian/RAICc}}, where we also provide the code to reproduce all of the simulation results in this paper. The sample sizes considered are $n\in\{40, 1000\}$, with number of predictors $p=n-1$, being close to the sample size. Predictors exhibit moderate correlation with each other of an AR(1) type (see the Online Supplemental material for further details), and the strength of the overall regression is characterized as either low (average $R^2$ on the set of true predictors roughly $20\%$) or high (average $R^2$ on the set of true predictors roughly $90\%$). The true model is either sparse (with six nonzero slopes) or dense (with $p$ nonzero slopes exhibiting diminishing strengths of coefficients; \citealp{Taddy2017}). The design matrix $X$ is random. In each replication, we generate a matrix $X$ such that the rows $x_i$ ($i=1,\cdots,n$) are drawn from a $p$-dimensional multivariate normal distribution with mean zero and covariance matrix $\Sigma_0$, and we draw the response $y$ from the conditional distribution of $y|X$ based on \eqref{eq:truemodel}. The entire process is repeated $1000$ times. 

We consider the following metrics to evaluate the fit. The values of each criterion over all of the simulation runs are plotted using side-by-side boxplots, with the average value over the simulation runs given below the boxplot for the corresponding criterion.
\begin{itemize}
  \item Root mean squared error for random-X:
  \begin{equation*}
    \text{RMSER} = \sqrt{ E_{X^n} \lVert X^n\hat\beta-X^n\beta_0 \rVert_2^2 } =  \sqrt{ (\hat{\beta}-\beta_0)^T \Sigma_0 (\hat{\beta}-\beta_0) }.
  \end{equation*} 

  \item KL discrepancy for random-X \eqref{eq:KLR} in the log scale (denoted as logKLR).

  \item Size of the subset selected for variable selection problem, and number of restrictions in the selected model for general restriction problem.
\end{itemize}

\begin{figure}[!ht]
  \centering
  \includegraphics[width=\textwidth]{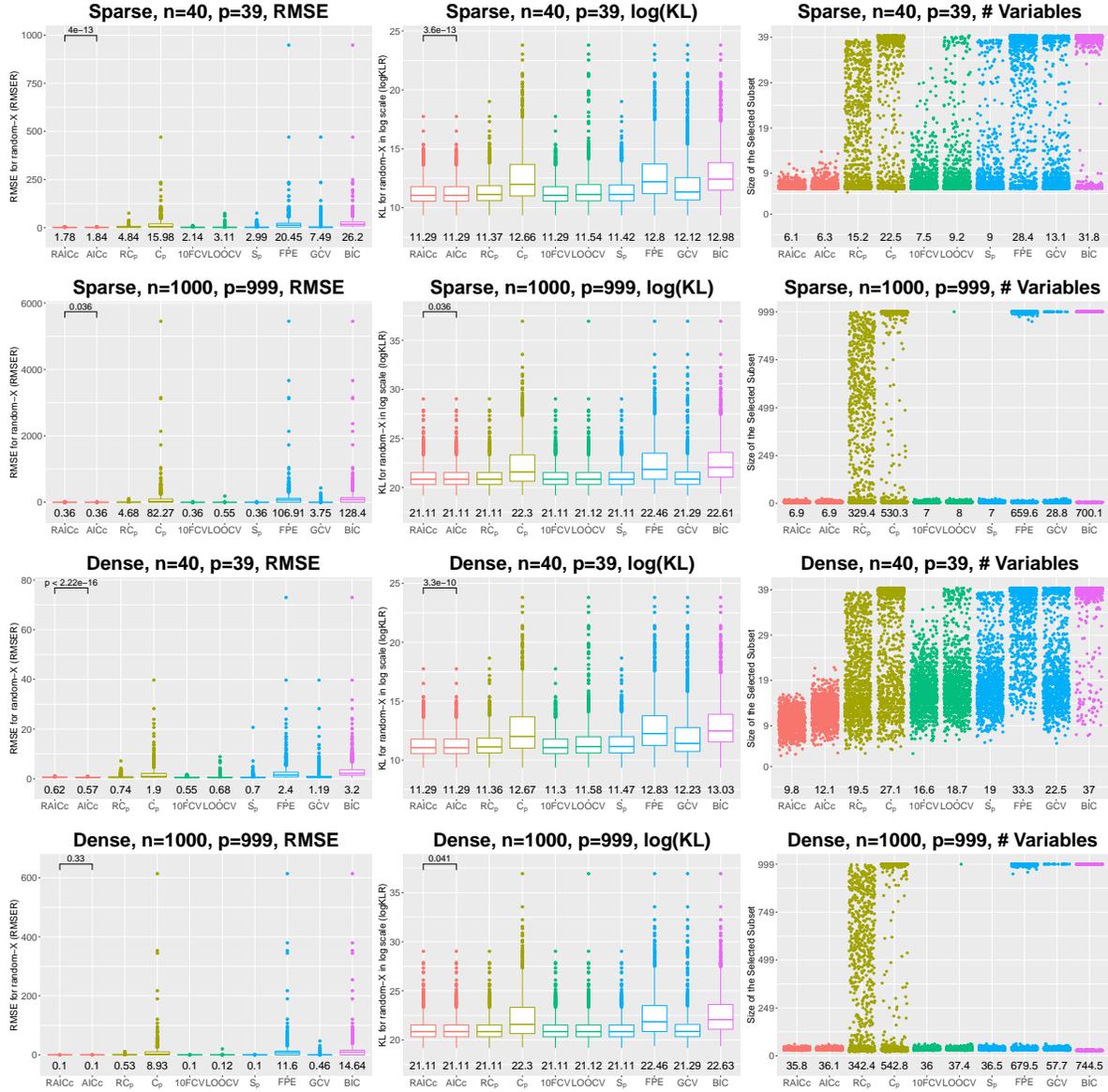}
  \caption{Results of simulations for variable selection. Random-X, high signal and $\rho=0.5$. The Sparse and Dense models correspond to the VS-Ex2 and VS-Ex3 configurations (details are given in the Online Supplemental Material). The first column refers to RMSE, the second column corresponds to KL discrepancy (in log scale), and the third column gives the number of variables in the selected model with nonzero slopes, jittered horizontally and vertically, so the number of models with that number of nonzero slopes can be ascertained more easily. The mean values of the evaluation metrics for each criterion are presented at the bottom of each graph. The p-values of the Wilcoxon signed-rank test (paired and two-sided) for comparing RAICc and AICc are also presented.}
  \label{fig:subsetselection_randomx_hsnr_largep}
\end{figure}

\begin{figure}[!ht]
  \centering
  \includegraphics[width=\textwidth]{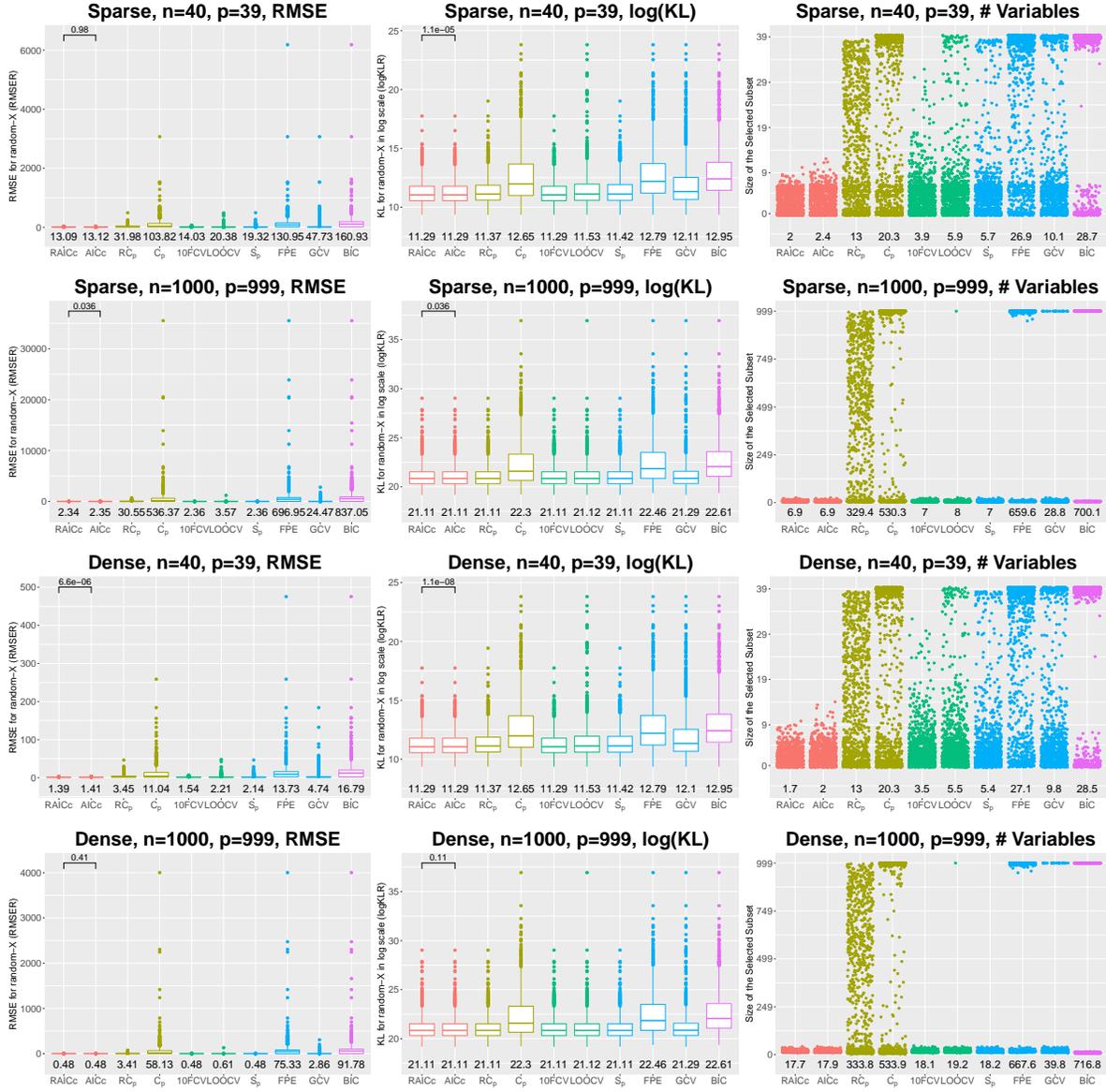}
  \caption{Results of simulations for variable selection. Random-X, low signal and $\rho=0.5$. }
  \label{fig:subsetselection_randomx_lsnr_largep}
\end{figure}

The results are presented in Figures \ref{fig:subsetselection_randomx_hsnr_largep} and \ref{fig:subsetselection_randomx_lsnr_largep}. We find that RAICc provides the best predictive performance and the sparsest subset while rarely underfitting, compared to other information criteria designed for random-X, including RC$_p$ and S$_p$. The underperformance of RC$_p$ and S$_p$ is due to overfitting. S$_p$, as an estimate of the squared prediction error, is a product of an unlogged residual sum of squares and a penalty term that is increasing in $k$. This results in higher variability in S$_p$ for models that overfit, thereby potentially increasing the chances for spurious minima of these criteria at models that drastically overfit. In RAICc, on the other hand, the residual sum of squares is logged, thereby stabilizing the variance and avoiding the problem. RC$_p$ drastically overfits in all scenarios, reflecting the price of estimating $\sigma_0^2$ using the full model, especially when $p$ is close to $n$. S$_p$, on the other hand, estimates $\sigma_0^2$ using the candidate model, which mitigates the problem. Nevertheless, S$_p$ also can sometimes strongly overfit, but only when $n$ is small. Even for large $n$, S$_p$ selects slightly larger subsets on average than does RAICc. 

We also note that information criteria designed for random-X generally perform better than their counterparts for the fixed-X case. Both C$_p$ and FPE are largely outperformed by RC$_p$ and S$_p$, respectively. The advantage of RAICc over AICc is statistically significant in most scenarios, based on the Wilcoxon signed-rank test (the $p$-value for the test comparing the criteria for RAICc and AICc is given above the first two boxes in the first two columns of the table), but is not obvious in a practical sense. The only place that we see an advantage of AICc is for \{Dense model, $n=40$ and high signal\}. In this scenario, a model with many predictors with nonzero slopes can predict well, but that advantage disappears when there is a relatively weak signal, as in that situation the added noise from including predictors with small slopes cannot be overcome by a small error variance. 

We further note that choosing the appropriate family of information criteria (the KL-based AICc and RAICc) is more important than choosing the information criteria designed for the underlying designs of $X$. AICc, despite being designed for fixed-X, outperforms RC$_p$ and S$_p$, which are designed for random-X in all of the scenarios, in terms of both predictive performance and providing sparse results. The KL-based criteria have a clear advantage compared to the squared-error based criteria.

Finally, we note some other findings that have been discussed previously in the literature. Despite its apparently strong penalty, BIC often chooses the model using all predictors when $n$ is close to $p$, as discussed in \citet{hurvich1989regression} and \citet{baraud2009gaussian}. We also see that even though GCV has a similar penalty term as S$_p$, it is more likely to suffer from overfitting. Unlike S$_p$, GCV can sometimes drastically overfit even when $n$ is large. The overfitting problem of GCV was also observed in the context of smoothing by \citet{hurvich1998smoothing}. We further find that 10-fold CV performs better than LOOCV, the latter of which sometimes drastically overfits. The tendency of LOOCV to strongly overfit was noted by \citet{scott1987biased} and \citet{hall1991local} in the context of smoothing. \citet{zhang2015cross} showed that when applied as selection rules, the larger validation set used by 10-fold CV can better distinguish the candidate models than can LOOCV, and this results in a model with smaller predictive error. RAICc performs better than 10-fold CV for small $n$, and performs similarly for large $n$. Computationally, 10-fold CV is ten times more expensive compared to RAICc, and since the split of validation samples is random, 10-fold CV can select different subsets if applied multiple times on the same dataset; that is, the result of 10-fold CV is not reproducible. The fact that LOOCV provides better estimate of the test error while being outperformed by 10-fold CV, further emphasizes the difference between the goal of providing the best estimate of the test error, and the goal of selecting the models with the best predictive performance. Clearly, KL-based criteria (AICc and RAICc) bridge the gap between the two goals more effectively than the squared-error based criteria (including cross-validation).

In a related study by \citet{leeb2008evaluation}, the author found that S$_p$ and GCV outperform AICc under random-X, but those results are not directly comparable to ours. That paper did not consider the case where $p$ is extremely close to $n$, which is the scenario that most separates the performances of the different criteria. 

We next consider the general restriction problem. We take $\beta_0 = [2,2,2,1,1,1]^T$, $n\in\{10,40\}$, moderate correlations between the predictors, and either high or low signal levels. The candidate models are constructed in the following way. We consider a set of restrictions: $\beta_1=\beta_4$, $\beta_1=2\beta_2$, $\beta_1=\beta_2$, $\beta_2=\beta_3$, $\beta_4=\beta_5$, $\beta_5=\beta_6$, where the last four restrictions hold for our choice of $\beta_0$. We then consider all of the possible subsets of the six restrictions, resulting in $64$ candidate models in total. The detailed configurations and complete results for this and other examples of the general restriction problem ($54$ scenarios in total) are given in the Online Supplemental Material. 

\begin{figure}[!ht]
  \centering
  \includegraphics[width=\textwidth]{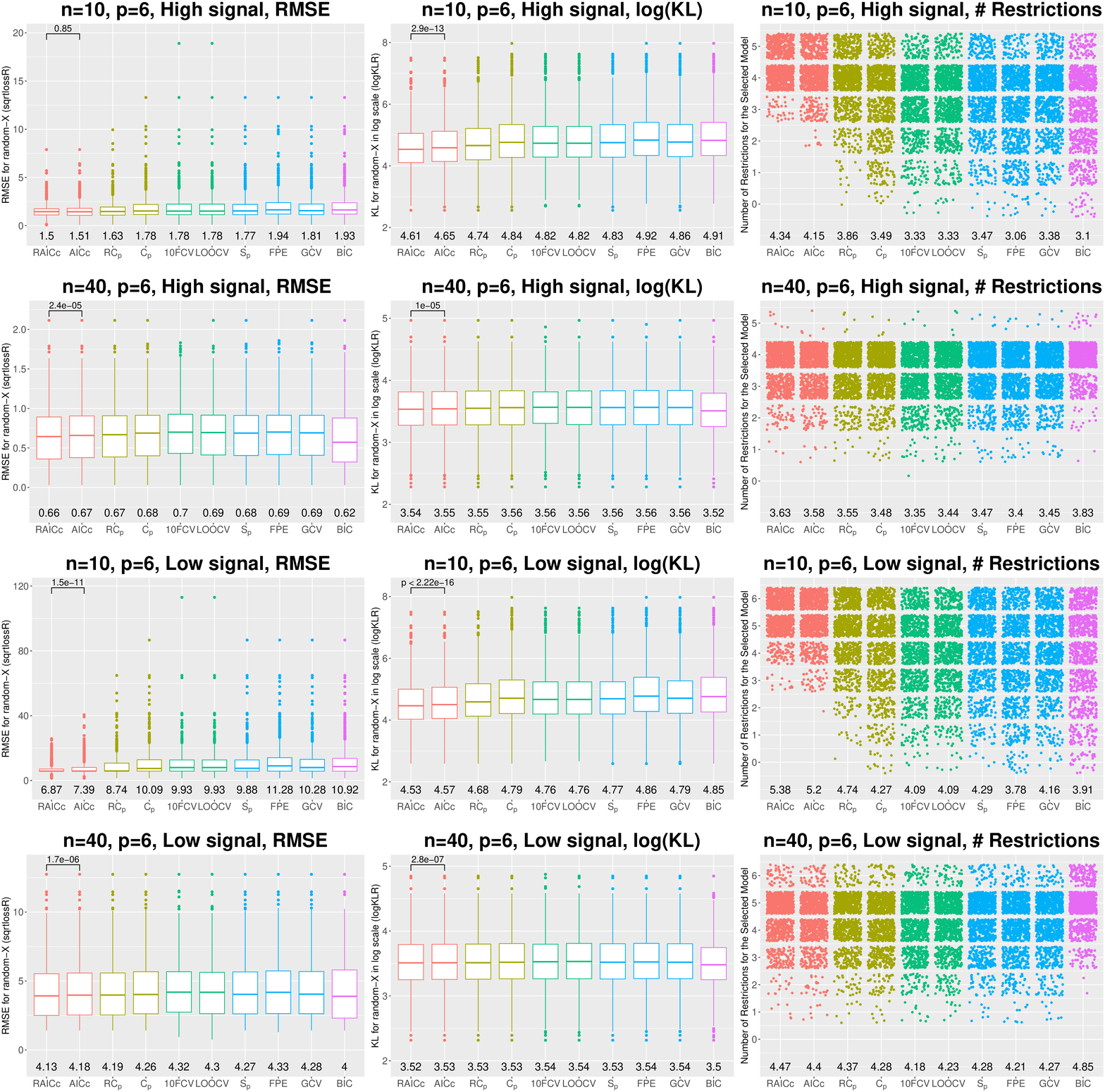}
  \caption{Results of simulations for general restrictions. Random-X, $\rho=0.5$. The configuration of the model is GR-Ex1 (details can be found in the Online Supplemental Material). Third column gives the number of restrictions in the selected models, jittered horizontally and vertically. }
  \label{fig:generalrestriction_randomx}
\end{figure}

We see from Figure \ref{fig:generalrestriction_randomx} that differences in performance between the criteria are less dramatic. This is not surprising, since for these models the number of parameters never approaches the sample size. Still, RAICc is consistently the best selection rule for small sample size $n$, and it is second-best for large $n$, where it is outperformed by BIC (note that BIC has a strong tendency to select too few restrictions when the sample is small, which corresponds to overfitting in the variable selection context). We also note an advantage of RAICc over AICc, with AICc having a stronger tendency to select too few restrictions.

Finally, we extend the general restriction example by including restrictions that force additional predictors to have zero coefficients (as in the variable selection problem). Besides the six restrictions specified, we also consider $\beta_i=0$ for $i=7,\cdots,p$ resulting in $p$ possible restrictions in total. The candidate models are formulated by excluding the restrictions in a nested fashion. We start from the model including all $p$ restrictions (corresponding to the null model), and the next model includes the $p-1$ restrictions except the first one $\beta_1=\beta_4$. The process is repeated until all restrictions are excluded (the full model including all predictors with arbitrary slopes) resulting in $p+1$ candidate models in total. The true coefficient vector is the same as that used in Figure \ref{fig:generalrestriction_randomx}, implying that the correct number of restrictions is $p-2$. We present the detailed configurations and complete results for this and other examples ($243$ scenarios in total) in the Online Supplemental Material. 

\begin{figure}[!ht]
  \centering
  \includegraphics[width=\textwidth]{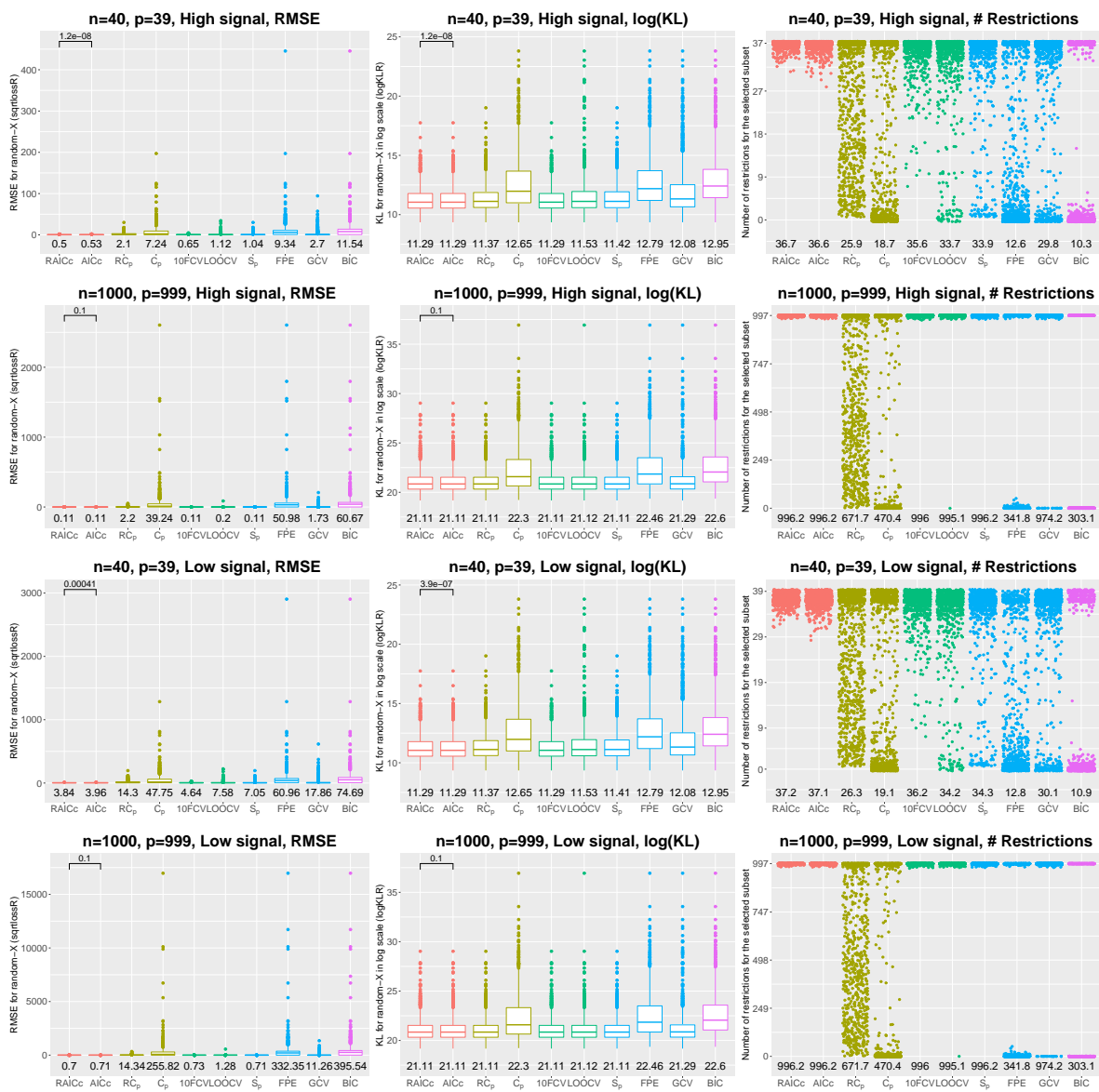}
  \caption{Results of simulations for general restrictions. Random-X, $\rho=0.5$. The configuration of the model is GR-Ex4 (details can be found in the Online Supplemental Material).}
  \label{fig:subsetgeneral_randomx}
\end{figure}

We see from Figure \ref{fig:subsetgeneral_randomx} that our findings for the variable selection problem also hold in this case. This is not surprising, since variable selection is just a special example of general restrictions, and in this scenario the set of candidate models includes ones where the number of parameters is close to the sample size. Thus, overall, RAICc and AICc are the best performers among all of the selectors. RAICc tends to provide the sparsest subset (or select more restrictions), while rarely underfitting, having a slight advantage over AICc in terms of predictive performance. 

\subsection{Fixed-X}
The simulation structure for random-X can also be applied to fixed-X. We only generate the design matrix $X$ once and draw $1000$ replications of the response vector $y$ from the conditional distribution of $y|X$ based on \eqref{eq:truemodel}. The evaluation metrics for fixed-X are as follows. The complete simulation results are given in the Online Supplemental Material. 
\begin{itemize}
  \item Root mean squared error for fixed-X:
  \begin{equation*}
    \text{RMSEF} = \sqrt{ \frac{1}{n}\lVert X\hat\beta-X\beta_0 \rVert_2^2 }.
  \end{equation*} 

  \item KL discrepancy for fixed-X \eqref{eq:KLF} in the log scale (denoted as logKLF).

  \item Size of the subset selected for variable selection problem, and number of restrictions in the selected model for general restriction problem.
\end{itemize}

\begin{figure}[!ht]
  \centering
  \includegraphics[width=\textwidth]{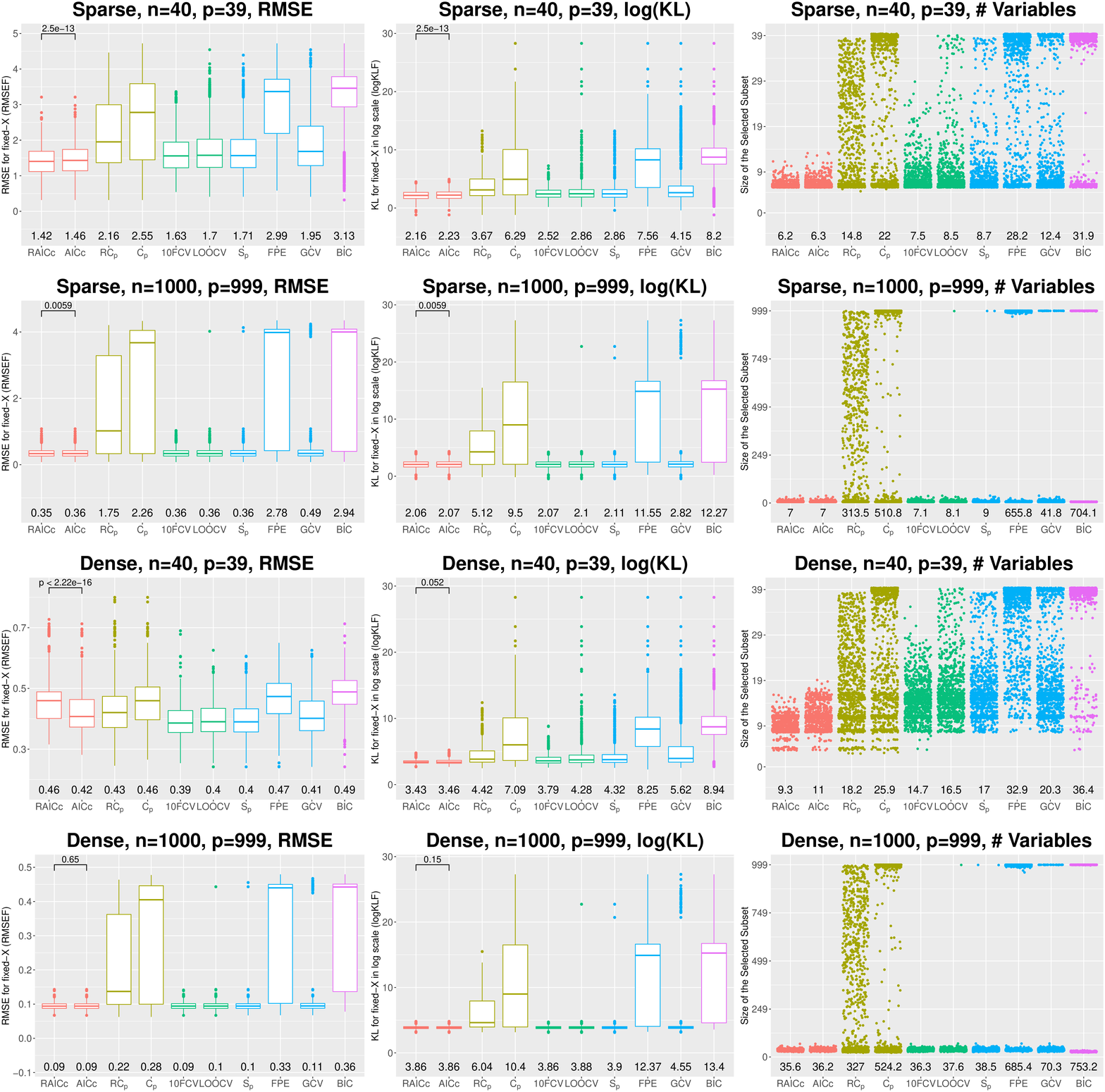}
  \caption{Results of simulations for variable selection. Fixed-X, high signal. The configurations are the same as in Figure \ref{fig:subsetselection_randomx_hsnr_largep}.}
  \label{fig:subsetselection_fixedx_hsnr_largep}
\end{figure}

\begin{figure}[!ht]
  \centering
  \includegraphics[width=\textwidth]{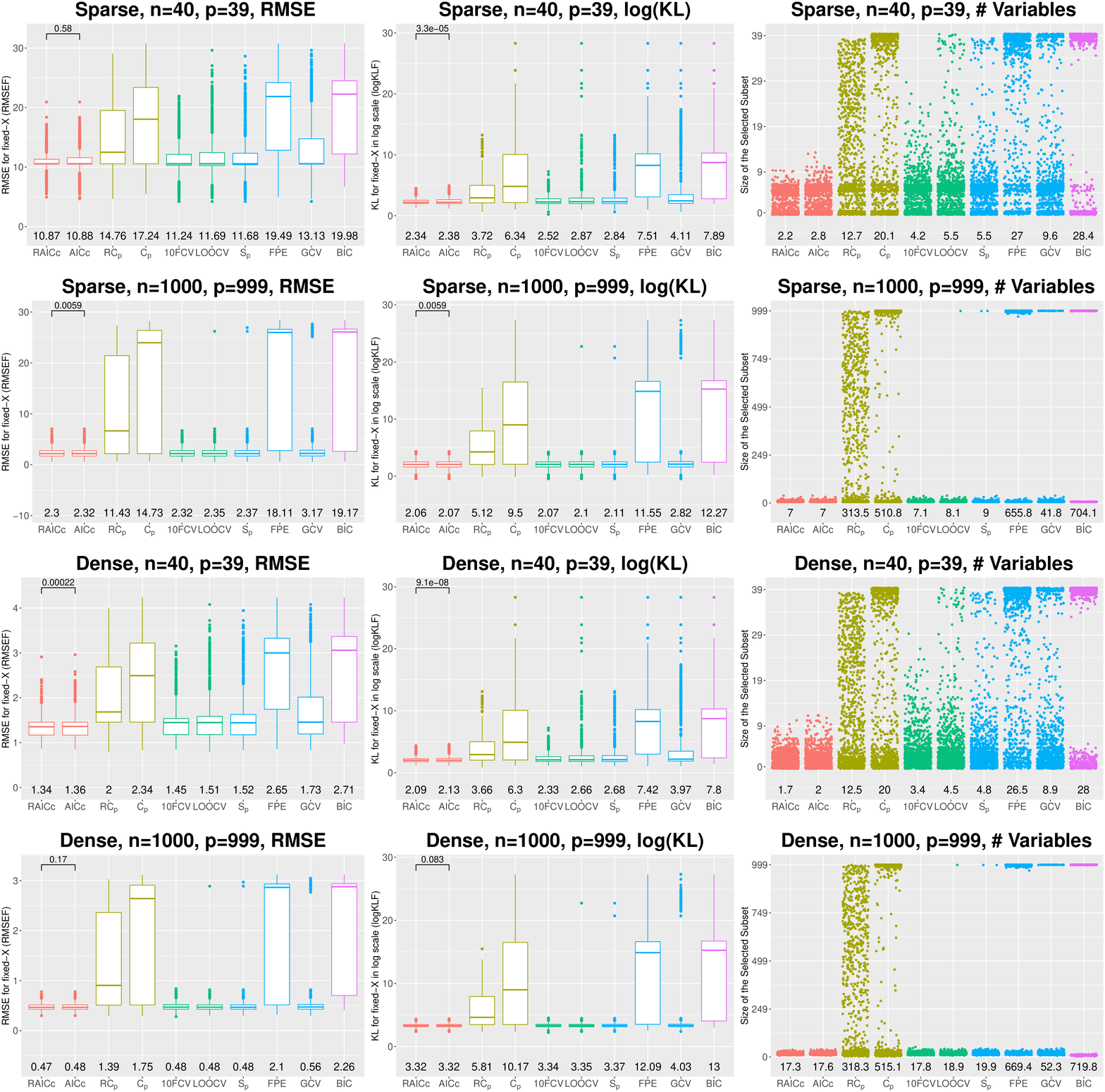}
  \caption{Results of simulations for variable selection. Fixed-X, low signal. The configurations are the same as in Figure \ref{fig:subsetselection_randomx_lsnr_largep}.}
  \label{fig:subsetselection_fixedx_lsnr_largep}
\end{figure}

\begin{figure}[!ht]
  \centering
  \includegraphics[width=\textwidth]{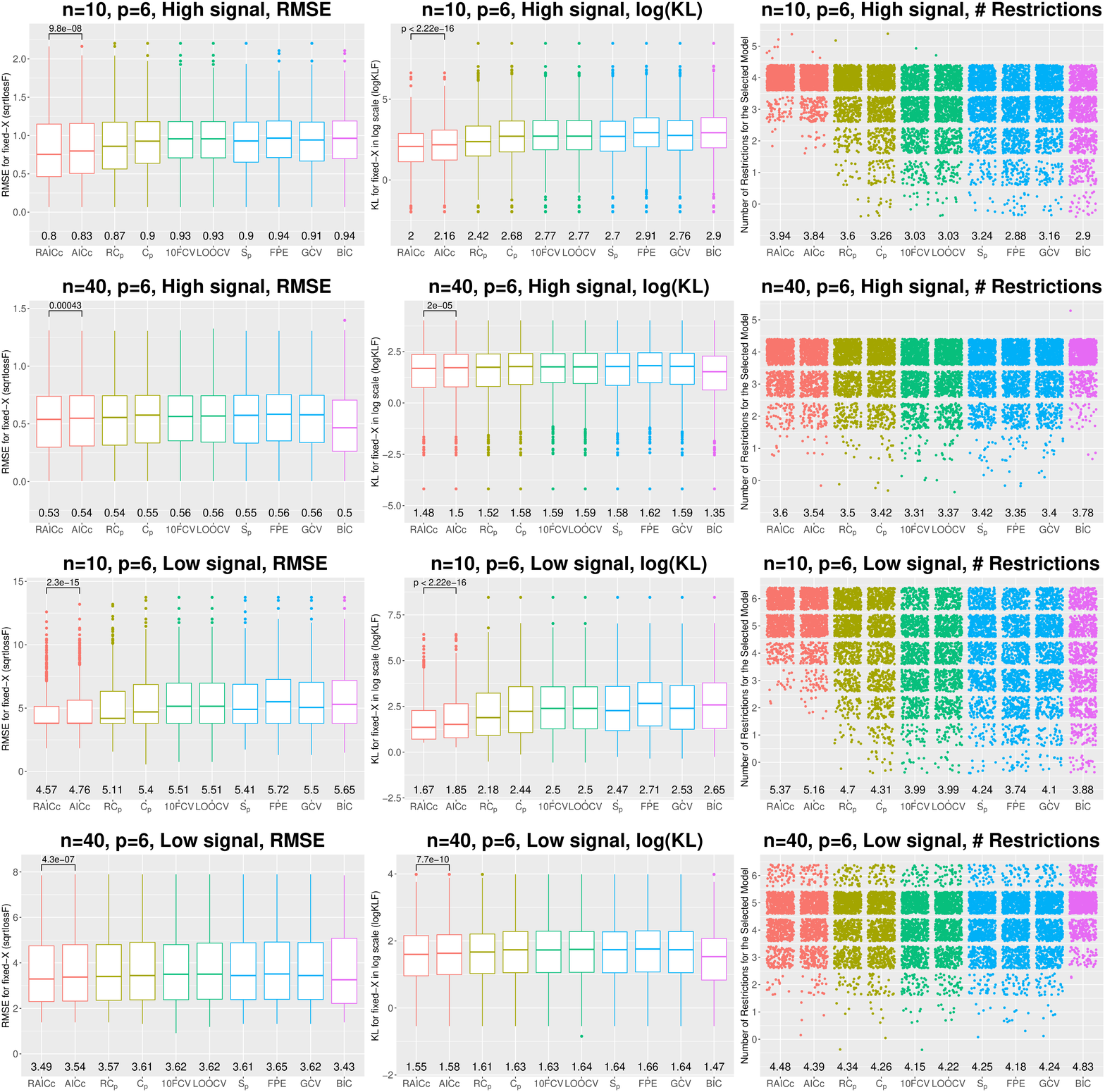}
  \caption{Results of simulations for general restrictions. Fixed-X, GR-Ex1, $\rho=0.5$. The configurations are the same as in Figure \ref{fig:generalrestriction_randomx}.}
  \label{fig:generalrestriction_fixedx}
\end{figure}

\begin{figure}[!ht]
  \centering
  \includegraphics[width=\textwidth]{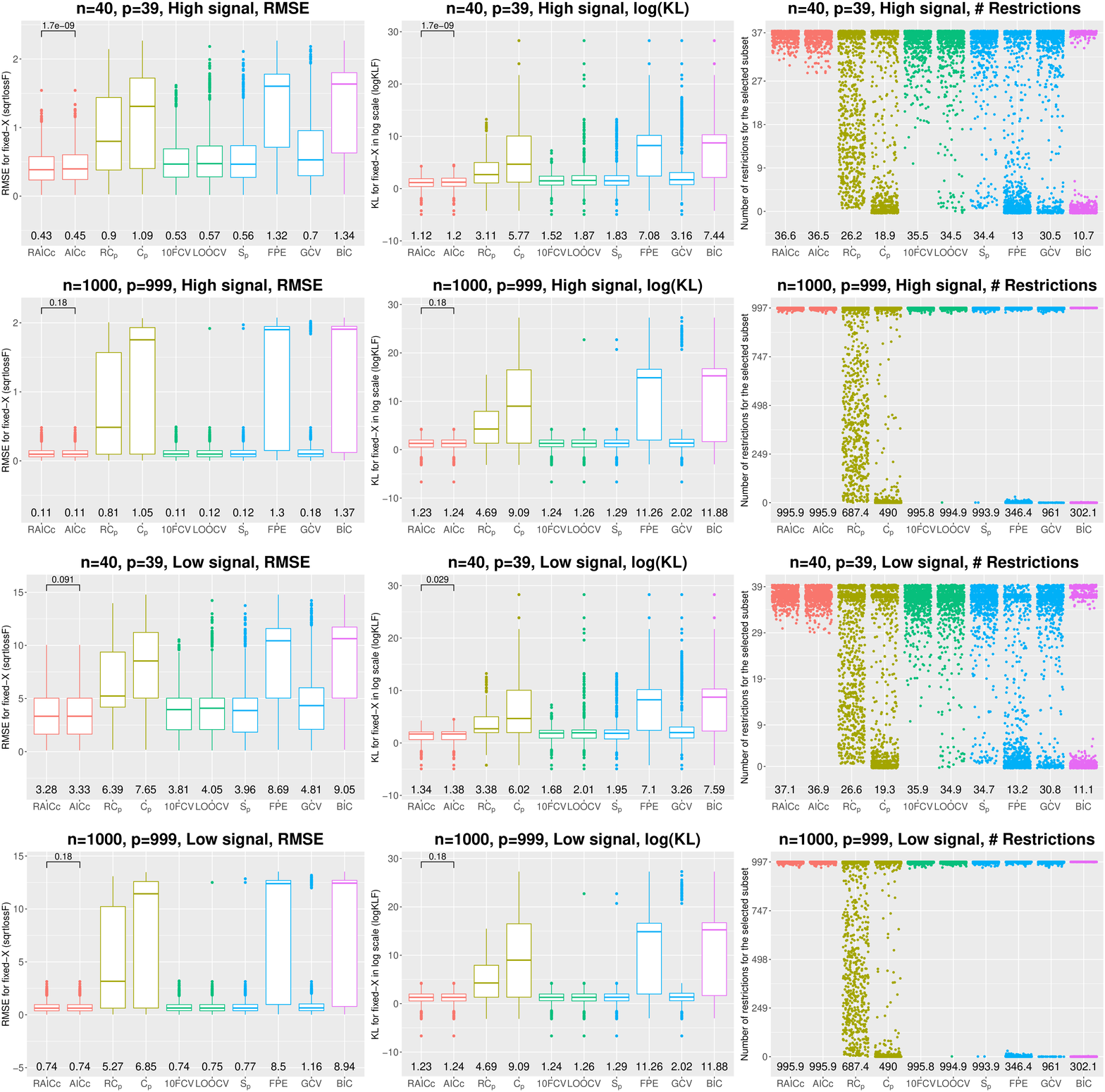}
  \caption{Results of simulations for general restrictions. Fixed-X, GR-Ex4, $\rho=0.5$. The configurations are the same as in Figure \ref{fig:subsetgeneral_randomx}.}
  \label{fig:subsetgeneral_fixedx}
\end{figure}

The patterns for the fixed-X scenario are similar to those for random-X, as can be seen in Figures \ref{fig:subsetselection_fixedx_hsnr_largep}, \ref{fig:subsetselection_fixedx_lsnr_largep}, \ref{fig:generalrestriction_fixedx} and \ref{fig:subsetgeneral_fixedx}. In some ways this is surprising, in that the random-X versions of the criteria still seem to outperform the fixed-X versions, even though that is not the scenario for which they are designed. This seems to be related to the tendency for the fixed-X versions to overfit (or choose too few restrictions) compared to their random-X counterparts, which apparently works against the goal of selecting the candidate with best predictive performance. Otherwise, the KL-based criteria (RAICc and AICc) noticeably outperform the other criteria in general, especially $\mbox{C}_p$ and FPE, particularly for small samples.

\section{Conclusion and future work}
\label{sec:conclusion}
In this paper, the use of information criteria to compare regression models under general linear restrictions for both fixed and random predictors is discussed. It is shown that general versions for KL-based discrepancy (AICc and RAICc, respectively) and squared error-based discrepancy (C$_p$, FPE, RC$_p$ and S$_p$, respectively) can be formulated as effectively unbiased estimators of the test error (up to some terms that are free of the linear restrictions and hence are irrelevant when comparing criteria for different models). Model comparison based on the KL-based discrepancy measures is shown via simulations to be better-behaved than squared error-based discrepancies (including cross-validation) in selecting models with low predictive error and sparse subset.

The study of RAICc for variable selection in this paper focuses on OLS fits on pre-fixed predictors (e.g. nested predictors based on their physical orders in $X$). The discussion can be extended to other fitting procedures where the predictors in each subset are decided in a data-dependent way. For instance, \citet{tian2019use} discussed using AICc for least-squares based subset selection methods, and extending those results to the random-X scenario is a topic for future work. 

Note also that only restrictions on the regression coefficients are considered here, corresponding to restrictions on the regression portion of the model. It is also possible that the data analyst could be interested in restrictions on the distributional parameters of the predictors (restricting the variances of some predictors to be equal to each other, for example, or restricting covariances to follow a specified pattern such as autoregressive of order $1$ or compound symmetry), and it would be interesting to try to generalize the criteria discussed here to that situation.

\clearpage
\bibliographystyle{chicago}

\bibliography{raicc.bib}

\beginsupplement
\appendix
\pagenumbering{arabic}
\begin{center}
\textbf{\large Supplemental Material \\
Selection of Regression Models under Linear Restrictions \\ for Fixed and Random Designs}

Sen Tian, Clifford M. Hurvich, Jeffrey S. Simonoff
\end{center}

This document provides theoretical details of the theorems and lemmas in the paper. The complete simulation results and the computer code to reproduce the results can be viewed online\footnote{\url{https://github.com/sentian/RAICc}}.

\section{Proof of Lemma \ref{thm:components_ekl_lr_fixedx}}
\begin{proof}
As is well known \citeponline[see, e.g.,][p.~122]{greene2003econometric},
\begin{equation}
  n\hat\sigma^2 = \lVert y-X\hat{\beta} \rVert_2^2 \sim \sigma_0^2 \chi^2(n-p+m),
  \label{eq:dist_rss}
\end{equation}
and by using Assumption \ref{assumption}
\begin{equation}
\begin{aligned}
E_y\left(\hat{\beta} - \beta_0 \right) &= 0,\\
\text{Cov}_y\left(\hat{\beta} - \beta_0 \right) &= E\left[\left(\hat{\beta} - \beta_0 \right)\left(\hat{\beta} - \beta_0 \right)^T \right]\\
&= \sigma_0^2 \left\{ (X^T X)^{-1} - (X^T X)^{-1}R^T\left[ R(X^T X)^{-1}R^T \right]^{-1} R(X^T X)^{-1} \right\}.
\end{aligned}
\label{eq:dist_betahat}
\end{equation}
From \eqref{eq:dist_rss}, $1/\hat{\sigma}^2$ follows an inverse $\chi^2$ distribution and we have
\begin{equation*}
n \sigma_0^2 E_y\left[ \frac{1}{\hat{\sigma}^2} \right] = n\frac{n}{n-p+m-2}.
\end{equation*}
From \eqref{eq:dist_betahat}, we have
\begin{equation*}
E_y  \left [ (\hat \beta-\beta_0)^T X^T X (\hat \beta-\beta_0) \right ] 
= \text{Tr} \left[ X^T X \cdot \text{Cov}_y \left( \hat{\beta} - \beta_0 \right) \right]
= \sigma_0^2 (p-m).
\end{equation*}
We next show that $\hat{\sigma}^2$ and $(\hat \beta-\beta_0)^T X^T X (\hat \beta-\beta_0)$ are independent. Define the idempotent matrix $H_R=(X^T X)^{-1} R^T \left[ R (X^T X)^{-1} R^T \right]^{-1} R$. Recall that two other idempotent matrices are defined as $H=X(X^T X)^{-1} X^T$ and $H_Q = X H_R (X^T X)^{-1} X^T$, respectively. We have
\begin{equation*}
\begin{aligned}
y-X\hat{\beta}^f &= (I-H)\epsilon,\\
X\hat{\beta}^f - X\hat{\beta} &= XH_R(\hat{\beta}^f - \beta_0) =H_Q \epsilon,\\
X\hat{\beta} - X\beta_0 &= X(I-H_R)(\hat{\beta}^f - \beta_0) = (H-H_Q) \epsilon,
\end{aligned}
\end{equation*}
where we use the fact that $\hat{\beta}^f - \beta_0 = (X^T X)^{-1}X^T \epsilon$. Also since $HH_Q=H_QH=H_Q$, any two of the three idempotent symmetric matrices $I-H$, $H_Q$ and $H-H_Q$ have product zero. Then by Craig's Theorem \citeponline{craig1943note} on the independence of two quadratic forms in a normal vector, 
\begin{equation*}
n\hat{\sigma}^2 = \lVert y-X\hat{\beta} \rVert_2^2 = \lVert y-X\hat{\beta}^f \rVert_2^2 + \lVert X\hat{\beta}^f - X\hat{\beta} \rVert_2^2 = \epsilon^T (I-H) \epsilon + \epsilon^T H_Q \epsilon
\end{equation*}
and
\begin{equation*}
\lVert X\hat{\beta} - X\beta_0 \rVert_2^2 = \epsilon^T (H-H_Q) \epsilon
\end{equation*}
are independent.
\end{proof}

\section{Proof of Theorem \ref{thm:EoptF_KL}}
\begin{proof}
By using Lemma \ref{thm:components_ekl_lr_fixedx}, the expected KL discrepancy can be derived as
\begin{equation*}
\begin{aligned}
E_y [\text{ErrF}_\text{KL} ] 
&= E_y \left\{ n \log (2\pi \hat\sigma^2) + \frac{1}{\hat\sigma^2}  (\hat\beta-\beta_0)^T X^T X (\hat\beta-\beta_0) + \frac{n\sigma_0^2}{\hat\sigma^2} \right\} \\
&= E_y \left [ n\log (2\pi \hat \sigma^2)\right ] +  (p-m) \frac{n}{n-p+m-2} + n \frac{n}{n-p+m-2}\\
&= E_y \left [ n\log (2\pi \hat \sigma^2) \right ] +  n \frac{n+p-m}{n-p+m-2}.
\end{aligned}
\end{equation*}
Recall that
\begin{equation*}
\text{errF}_\text{KL} = n\log(2\pi \hat\sigma^2) + n.
\end{equation*}
The expected optimism is then
\begin{equation*}
E_y(\text{optF}_\text{KL}) = E_y[\text{ErrF}_\text{KL} ]  -  E_y[\text{errF}_\text{KL} ] = n \frac{n+p-m}{n-p+m-2} - n.
\end{equation*}
\end{proof}

\section{Proof of Theorem \ref{thm:EoptF_SE}}
\begin{proof}
Using the expression of $\hat\beta$ \eqref{eq:betahat_sigmahatsq} and the definitions of $H$ and $H_Q$, we have
\begin{equation*}
\hat{\mu} = X\hat{\beta} = (H-H_Q)y + X(X^T X)^{-1} R^T \left[ R(X^TX)^{-1}R^T  \right]^{-1}r,
\end{equation*}
where the second term on the right-hand side is deterministic. Denote $h_i$ and ${h_Q}_i$ as the $i$-th rows of $H$ and $H_Q$, respectively. We then have
\begin{equation*}
\text{Cov}_y \left(\hat{\mu}_i, y_i \right) = \text{Cov}_y \left[ (h_i-{h_Q}_i)y, y_i  \right] = \text{Cov}_y \left[ (H_{ii}-{H_Q}_{ii})y_i, y_i \right] = \sigma_0^2 (H_{ii}-{H_Q}_{ii}).
\end{equation*}
Therefore, the covariance penalty \eqref{eq:EoptF_SE} can be derived as
\begin{equation*}
E_y (\text{optF}_\text{SE}) = 2 \sum_{i=1}^n \text{Cov}_y (\hat\mu_i,y_i) = 2 \sigma_0^2 \text{Tr}(H-H_Q) = 2 \sigma_0^2 (p-m).
\end{equation*}
\end{proof}

\section{Proof of Lemma \ref{thm:components_ekl_lr_randomx}}
\begin{proof}
Since $x_i$ are iid $\mathcal{N}(0,\Sigma_0)$, $X^T X \sim \mathcal{W}(\Sigma_0, n)$ and $ (X^T X)^{-1} \sim \mathcal{W}^{-1} (\Sigma_0^{-1}, n)$, where $\mathcal{W}$ and $\mathcal{W}^{-1}$ denotes a Wishart and an inverse Wishart distribution with $n$ degrees of freedom, respectively. We have $E(X^T X) = n\Sigma_0$ and $E((X^T X)^{-1}) = \Sigma_0^{-1} / (n-p-1)$. Hence,
\begin{equation*}
  E \left[ \text{Tr}(\hat \Sigma^{-1}\Sigma_0) \right] = E \left[ \text{Tr} (n (X^T X)^{-1} \Sigma_0) \right] = n \text{Tr}\left[ E\left( (X^T X)^{-1} \right) \Sigma_0\right] = \frac{np}{n-p-1}.
\end{equation*}
Define $H_S = X(X^T X)^{-1}(I-H_R)^T \Sigma_0 (I-H_R) (X^T X)^{-1} X^T$. Conditionally on $X$, the random variable $\hat{\sigma}^2$ and 
\begin{equation*}
(\hat{\beta}-\beta_0)^T \Sigma_0 (\hat{\beta}-\beta_0) = \epsilon^T H_S \epsilon
\end{equation*}
are independent by Craig's Theorem, since $H_S$ is symmetric and $H_S(I-H+H_Q)=0$. 
In order to calculate $\displaystyle E_{X,y}  \left [ (\hat \beta-\beta_0)^T \Sigma_0 (\hat \beta-\beta_0) \right ]$, we transform the original basis of the problem. Denote $\tilde{R} = \big(\begin{smallmatrix}
  R\\
  R^c
\end{smallmatrix}\big)$, a ($p \times p$) matrix, where the rows of $R^c$ span the orthogonal complement of the row space of $R$. Hence $\tilde{R}$ has full rank. The true model now becomes
\begin{equation*}
y = X \beta_0 + \epsilon = \tilde{X} \tilde{\beta}_0 + \epsilon,
\end{equation*}
where $\tilde{X} = X \tilde{R}^T$, $\tilde{\beta}_0 = \tilde{R}^{T^{-1}} \beta_0$. Denote $\tilde{M} = R \tilde{R}^T$. Assumption \ref{assumption} indicates that the true model in the new basis satisfies $\tilde{M} \tilde{\beta}_0 = r$. The approximating model is
\begin{equation*}
y = X \beta + u = \tilde{X} \tilde{\beta} + u,
\end{equation*}
with restrictions $\tilde{M} \tilde{\beta} = r$ where $\tilde{\beta} = \tilde{R}^{T^{-1}} \beta$. Denote $\hat{\tilde{\beta}}^f = (\tilde{X}^T \tilde{X})^{-1}\tilde{X}^Ty$ as the OLS estimator in the regression of $y$ on $\tilde{X}$. The restricted MLE is then
\begin{equation*}
\hat{\tilde{\beta}} = \hat{\tilde{\beta}}^f - \left( \tilde{X}^T \tilde{X} \right)^{-1} \tilde{M}^T \left[ \tilde{M} (\tilde{X}^T \tilde{X})^{-1} \tilde{M}^T \right]^{-1} \left( \tilde{M} \hat{\tilde{\beta}}^f - r \right),
\end{equation*}
and it can be easily verified that $\hat{\tilde{\beta}} = \tilde{R}^{T^{-1}} \hat{\beta}$. Denote $\tilde{X}_m$ and $\tilde{X}_{p-m}$ as the matrices containing the first $m$ and last $p-m$ columns of $\tilde{X}$, respectively. Let $\hat{\tilde{\beta}}_{m}$ and $\hat{\tilde{\beta}}_{p-m}$ be column vectors consisting of the first $m$ and last $p-m$ entries in $\hat{\tilde{\beta}}$, respectively. Also let $\tilde{\beta}_{0,m}$ and $\tilde{\beta}_{0,p-m}$ be column vectors consisting of the first $m$ and last $p-m$ entries in $\tilde{\beta}_0$, respectively. By using the formula for the inverse of partitioned matrices and some algebra, it can be shown that (details are given in Supplemental Material Section \ref{sec:derivation_betahattilde})
\begin{equation}
\begin{aligned}
\hat{\tilde{\beta}}_m &= \tilde{r},\\
\hat{\tilde{\beta}}_{p-m} &= \left( \tilde{X}_{p-m}^T \tilde{X}_{p-m} \right)^{-1} \left( \tilde{X}_{p-m}^Ty - \tilde{X}_{p-m}^T\tilde{X}_{m}\tilde{r} \right),
\end{aligned}
\label{eq:betahat_tilde_partition}
\end{equation}
where $\tilde{r} = (R R^T)^{-1}r$. The restrictions $\tilde{M}\tilde{\beta}_0=r$ results in $\tilde{\beta}_{0,m} = \tilde{r}$. We then have 
\begin{equation*}
\hat{\tilde{\beta}}_{p-m} - \tilde{\beta}_{0,p-m} = \left( \tilde{X}_{p-m}^T \tilde{X}_{p-m} \right)^{-1} \tilde{X}_{p-m} ^T \epsilon,
\end{equation*}
and therefore 
\begin{equation*}
\hat{\tilde{\beta}}_{p-m} - \tilde{\beta}_{0,p-m} \big | \tilde{X} \sim \mathcal{N} \left(0, \sigma_0^2 \left( \tilde{X}_{p-m}^T \tilde{X}_{p-m} \right)^{-1} \right).
\end{equation*}
We also note that $\tilde{X}_{p-m} = X {R^c}^T$, and hence the rows $\tilde{x}_{p-m, i}$ of $\tilde{X}_{p-m}$, are independent and satisfy $\tilde{x}_{p-m, i} \sim \mathcal{N} \left( 0, R^c \Sigma_0 {R^c}^T \right)$. We then have that $\left( \tilde{X}_{p-m}^T \tilde{X}_{p-m} \right)^{-1}$ follows the inverse Wishart distribution $W^{-1}( R^c \Sigma_0 {R^c}^T, n)$. The expectation of the quadratic form can be derived as
\begin{equation*}
\begin{aligned}
E_{X,y}  \left [ (\hat \beta-\beta_0)^T \Sigma_0 (\hat \beta-\beta_0) \right ] 
&=  E_{\tilde{X},y}  \left [ \left( \hat{\tilde{\beta}} - \tilde{\beta}_0 \right)^T \tilde{R} \Sigma_0 \tilde{R}^T \left( \hat{\tilde{\beta}} - \tilde{\beta}_0 \right) \right ]\\
&=  E_{\tilde{X}} \left\{ E  \left [ \left( \hat{\tilde{\beta}}_{p-m} - \tilde{\beta}_{0,p-m} \right)^T R^c \Sigma_0 {R^c}^T \left(\hat{\tilde{\beta}}_{p-m} - \tilde{\beta}_{0,p-m} \right) \Big| \tilde{X} \right ]  \right\}\\
&= \sigma_0^2 \text{Tr} \left\{ R^c \Sigma_0 {R^c}^T E \left[ \left( \tilde{X}_{p-m}^T \tilde{X}_{p-m} \right)^{-1} \right] \right\} \\
&= \sigma_0^2 \frac{p-m}{n-p+m-1}.
\end{aligned}
\end{equation*}

\end{proof}

\section{Proof of Theorem \ref{thm:EoptR_KL}}
\begin{proof}
The expected KL can be derived as
\begin{equation*}
\begin{aligned}
&E_{X,y} (\text{ErrR}_\text{KL}) \\
&= E_{X,y} \left [ n \log (2\pi \hat\sigma^2) + \frac{n}{\hat\sigma^2}  (\hat\beta-\beta_0)^T \Sigma_0 (\hat\beta-\beta_0) + \frac{n\sigma_0^2}{\hat\sigma^2} \right ] + E \left [np \log(2\pi) + n \log |\hat\Sigma| + n \text{Tr}(\hat\Sigma^{-1}\Sigma_{0}) \right] \\
&= E_{X,y} \left [ n \log (2\pi \hat\sigma^2) \right ] + E_X \left[E\left(\frac{n}{\hat\sigma^2} \big | X \right)  E\left((\hat\beta-\beta_0)^T \Sigma_0 (\hat\beta-\beta_0) \big| X \right) + E\left(\frac{n\sigma_0^2}{\hat\sigma^2} \big| X \right) \right] + \\
 & \qquad E \left [np \log(2\pi) + n \log |\hat\Sigma| + n \text{Tr}(\hat\Sigma^{-1}\Sigma_{0}) \right] \\
&= E_{X,y} \left [ n\log (2\pi \hat \sigma^2) \right ] +  n \frac{n(n-1)}{(n-p+m-2)(n-p+m-1)} + E \left [ n\log |\hat \Sigma| \right ] + np\log(2\pi) + n \frac{np}{n-p-1},
\end{aligned}
\end{equation*}
where the second equality is based on Lemma \ref{thm:components_ekl_lr_randomx} for the independence of $\hat\sigma^2$ and $(\hat\beta-\beta_0)^T \Sigma_0 (\hat\beta-\beta_0)$ conditionally on $X$, and in the last equality we use results from Lemmas \ref{thm:components_ekl_lr_fixedx} and \ref{thm:components_ekl_lr_randomx}. Since the training error is
\begin{equation*}
\text{errR}_\text{KL} = -2\log L(\hat\beta,\hat\sigma^2,\hat\Sigma|X,y) = \left [ n \log (2\pi \hat\sigma^2) + n \right ] + \left [np \log(2\pi) + n \log |\hat\Sigma| + np \right ],
\end{equation*}
the expected optimism can be obtained as
\begin{equation*}
\begin{aligned}
E_{X,y}(\text{optR}_\text{KL}) 
&= E_{X,y}[\text{ErrR}_\text{KL} ]  -  E_{X,y}[\text{errR}_\text{KL} ] \\
&= n \frac{n(n-1)}{(n-p+m-2)(n-p+m-1)} + n \frac{np}{n-p-1} - n(p+1).
\end{aligned}
\end{equation*}
\end{proof}

\section{Proof of Theorem \ref{thm:EoptR_SE}}
\begin{proof}

We first note from Theorem \ref{thm:EoptF_SE} that
\begin{equation*}
E_{X,y}(\text{optF}_\text{SE}) = E \left[ E(\text{optF}_\text{SE} | X ) \right] = 2 \sigma_0^2(p-m).
\end{equation*}
Based on formula 6 and proposition 1 in \citetonline{rosset2020fixed}, the expected optimism can be decomposed into 
\begin{equation*}
E_{X,y}(\text{optR}_\text{SE}) = E_{X,y}(\text{optF}_\text{SE}) + B^+ + V^+ = 2 \sigma_0^2(p-m) +  B^+ + V^+,
\end{equation*}
where $B^+$ and $V^+$ are the excess bias and excess variance of the fit. In particular, the excess bias is defined as
\begin{equation*}
B^+ = E_{X,X^{(n)}} \big\lVert E( X^{(n)} \hat{\beta} \big | X, X^{(n)} ) - X^{(n)} \beta_0 \big\rVert_2^2 - E_X \big\lVert E( X \hat{\beta} \big | X ) - X \beta_0 \big\rVert_2^2.
\end{equation*}
Because of Assumption \ref{assumption} that the true model satisfies the restrictions, it follows that $\hat{\beta}$ is unbiased, and hence it is easy to see that $B^+=0$. Next, $V^+$ is defined as
\begin{equation*}
V^+ = E_{X,X^{(n)}} \left\{ \text{Tr} \left[ \text{Cov}\left( X^{(n)}\hat{\beta} \big | X, X^{(n)} \right) \right] \right\} - E_X \left\{ \text{Tr} \left[ \text{Cov}\left( X\hat{\beta} \big | X \right)  \right] \right\}.
\end{equation*}
The second term on the right-hand side is
\begin{equation*}
E_X \left\{  \text{Tr} \left[ \text{Cov} \left( X\hat{\beta} \big | X \right)  \right] \right\} = E_X \left\{ \text{Tr} \left[ \text{Cov} \left( (H-H_Q)y \big | X \right)  \right] \right\} = E\left\{  \sigma_0^2 \text{Tr} \left( H-H_Q  \right) \right\}= \sigma_0^2 (p-m).
\end{equation*}
The first term on the right-hand side is
\begin{equation*}
\begin{aligned}
& E_{X,X^{(n)}} \text{Tr} \left[ \text{Cov}\left( X^{(n)}\hat{\beta} \big | X, X^{(n)} \right) \right] \\
&= E_{X,X^{(n)}} \text{Tr} \left[ \text{Cov} \left( X^{(n)} (\hat{\beta} -\beta_0) \big | X, X^{(n)} \right) \right] \\
&= \text{Tr} \left\{ E_X\left[ \text{Cov} \left( \hat{\beta} -\beta_0 \big | X \right) \right] E \left ({X^{(n)}}^T X^{(n)}) \right] \right\}\\
&= n E_X \left\{ \text{Tr} \left[ \Sigma_0 \text{Cov} \left( \hat{\beta} - \beta_0  \big | X \right) \right] \right\}\\
&= n E_X \left\{ E\left[ \left(\hat{\beta} - \beta_0 \right)^T \Sigma_0 \left(\hat{\beta} - \beta_0 \right) \big| X \right] \right\} \\
&= n E_{X,y} \left[ \left(\hat{\beta} - \beta_0 \right)^T \Sigma_0 \left(\hat{\beta} - \beta_0 \right) \right]\\
&= n \sigma_0^2 \frac{p-m}{n-p+m-1},
\end{aligned}
\end{equation*}
where in the third equality we use the independence and identical distribution of $X$ and $X_0$, and $E(X^T X) = n\Sigma_0$, while in the last equality we use the result in Lemma \ref{thm:components_ekl_lr_randomx}. Combining the results together, we have
\begin{equation*}
V^+ = n\sigma_0^2 \frac{p-m}{n-p+m-1} - \sigma_0^2 (p-m) = \sigma_0^2\frac{(p-m)(p-m+1)}{n-p+m-1},
\end{equation*}
and 
\begin{equation*}
E_{X,y}(\text{optR}_\text{SE}) = 2\sigma_0^2 (p-m) + \sigma_0^2\frac{(p-m)(p-m+1)}{n-p+m-1}.
\end{equation*}

\end{proof}

\section{Derivation of the expression of \texorpdfstring{$\hat{\tilde{\beta}}$}{} in \eqref{eq:betahat_tilde_partition} }
\label{sec:derivation_betahattilde}

Denote $\tilde{H}_m = \tilde{X}_m \left(\tilde{X}_m^T \tilde{X}_m \right)^{-1} \tilde{X}_m^T$ and $\tilde{H}_{p-m} = \tilde{X}_{p-m} \left(\tilde{X}_{p-m}^T \tilde{X}_{p-m}\right)^{-1} \tilde{X}_{p-m}^T$. Then the partitioned form of the matrix $\tilde{X}^T \tilde{X}$ is given by
\begin{equation*}
\begin{aligned}
& \left(\tilde{X}^T \tilde{X}\right)^{-1} = 
\begin{bmatrix}
\tilde{X}_m^T \tilde{X}_m & \tilde{X}_m^T \tilde{X}_{p-m} \\
\tilde{X}_{p-m}^T \tilde{X}_m & \tilde{X}_{p-m}^T \tilde{X}_{p-m}
\end{bmatrix}^{-1} \\
&=
\begin{bmatrix}
\left[\tilde{X}_m^T (I-\tilde{H}_{p-m}) \tilde{X}_m \right]^{-1} & - \left[\tilde{X}_m^T (I-\tilde{H}_{p-m}) \tilde{X}_m \right]^{-1} \tilde{X}_m^T \tilde{X}_{p-m} \left(\tilde{X}_{p-m}^T \tilde{X}_{p-m}\right)^{-1}\\
- \left[\tilde{X}_{p-m}^T \left(I-\tilde{H}_m\right) \tilde{X}_{p-m} \right]^{-1} \tilde{X}_{p-m}^T \tilde{X}_m \left(\tilde{X}_m^T \tilde{X}_m \right)^{-1} & \left[\tilde{X}_{p-m}^T \left(I-\tilde{H}_m\right) \tilde{X}_{p-m} \right]^{-1}
\end{bmatrix},
\end{aligned}
\end{equation*}
and the partitioned form of $\hat{\tilde{\beta}}^f$ is given by
\begin{equation*}
\begin{aligned}
\hat{\tilde{\beta}}^f &= \left(\tilde{X}^T \tilde{X}\right)^{-1} \tilde{X}^T y \\
&= 
\begin{bmatrix}
\left[\tilde{X}_m^T \left(I-\tilde{H}_{p-m}\right) \tilde{X}_m \right]^{-1} \left( \tilde{X}_m^T y - \tilde{X}_m^T \tilde{H}_{p-m} y \right) \\
\left[\tilde{X}_{p-m}^T \left(I-\tilde{H}_m\right) \tilde{X}_{p-m} \right]^{-1} \left( \tilde{X}_{p-m}^T y - \tilde{X}_{p-m}^T \tilde{H}_m y \right)
\end{bmatrix}.
\end{aligned}
\end{equation*}
We also have 
\begin{equation*}
\begin{aligned}
& \left[\tilde{X}_{p-m}^T \left(I-\tilde{H}_m\right) \tilde{X}_{p-m} \right]^{-1} \tilde{X}_{p-m}^T \tilde{H}_m \left(I_n - \tilde{H}_{p-m}\right)\tilde{X}_m \\
&= \left[\tilde{X}_{p-m}^T \left(I-\tilde{H}_m\right) \tilde{X}_{p-m} \right]^{-1} \left( \tilde{X}_{p-m}^T \tilde{X}_m - \tilde{X}_{p-m}^T \tilde{H}_m\tilde{H}_{p-m}\tilde{X}_m \right)  \\
&= \left[\tilde{X}_{p-m}^T \left(I-\tilde{H}_m\right) \tilde{X}_{p-m} \right]^{-1} \tilde{X}_{p-m}^T \left(I-\tilde{H}_m\right) \tilde{X}_{p-m} \left(\tilde{X}_{p-m}^T \tilde{X}_{p-m} \right)^{-1} \tilde{X}_{p-m}^T \tilde{X}_m  \\
&= \left(\tilde{X}_{p-m}^T \tilde{X}_{p-m} \right)^{-1} \tilde{X}_{p-m}^T \tilde{X}_m.
\end{aligned}
\end{equation*}
Using this property and $\tilde{M} = R \tilde{R}^T = \big(\begin{smallmatrix}
  RR^T & 0
\end{smallmatrix}\big)$, we have
\begin{equation*}
\begin{aligned}
\left(\tilde{X}^T \tilde{X}\right)^{-1} \tilde{M}^T \left[ \tilde{M} \left( \tilde{X}^T \tilde{X} \right)^{-1} \tilde{M}^T \right]^{-1} &= 
\begin{bmatrix}
(RR^T)^{-1} \\
-\left(\tilde{X}_{p-m}^T \tilde{X}_{p-m} \right)^{-1} \tilde{X}_{p-m}^T \tilde{X}_m (RR^T)^{-1}
\end{bmatrix},
 \\ 
I_p - \left(\tilde{X}^T \tilde{X}\right)^{-1} \tilde{M}^T \left[ \tilde{M} \left( \tilde{X}^T \tilde{X} \right)^{-1} \tilde{M}^T \right]^{-1} \tilde{M} &= 
\begin{bmatrix}
0 & 0\\
\left(\tilde{X}_{p-m}^T \tilde{X}_{p-m} \right)^{-1} \tilde{X}_{p-m}^T \tilde{X}_m & I_{p-m}
\end{bmatrix}.
\end{aligned}
\end{equation*}
Therefore, \eqref{eq:betahat_tilde_partition} can be derived as
\begin{equation*}
\begin{aligned}
\hat{\tilde{\beta}} &= \left\{I_p - \left(\tilde{X}^T \tilde{X}\right)^{-1} \tilde{M}^T \left[ \tilde{M} \left( \tilde{X}^T \tilde{X} \right)^{-1} \tilde{M}^T \right]^{-1} \tilde{M} \right\}\hat{\tilde{\beta}}^f + \left\{\left(\tilde{X}^T \tilde{X}\right)^{-1} \tilde{M}^T \left[ \tilde{M} \left( \tilde{X}^T \tilde{X} \right)^{-1} \tilde{M}^T \right]^{-1} \right\}r\\
&=
\begin{bmatrix}
\tilde{r}\\
\left(\tilde{X}_{p-m}^T \tilde{X}_{p-m} \right)^{-1} \tilde{X}_{p-m}^T \left(y - \tilde{X}_m \tilde{r}\right)
\end{bmatrix}.
\end{aligned}
\end{equation*}

\bibliographystyleonline{chicago}
\bibliographyonline{raicc.bib}

\end{document}